\documentstyle[prl,aps,multicol,epsf,eqsecnum,epsfig]{revtex}

\begin{document}
\draft
\def \be #1{\begin{equation}\label{#1}}
\def \ee {\end{equation}}
\def \bea #1{\begin{eqnarray}\label{#1}}
\def \eea {\end{eqnarray}}
\def \OMIT #1{}
\def \rem #1{{\it #1}}
\def \REM #1{{\it #1}}
\def \Eq #1{Eq.~(\ref{#1})}
\def \Fig #1{Fig.~\ref{#1}}
\def \q {{\bf q}}
\def \F2 {FPL${}^2$ }
\def \Rs {\sf I\hskip-1.5pt R} 
\def \Zs {\mbox{\sf Z\hskip-5pt Z}} 
\def \Cs {\rm C\!\!\!I\:}
\def \bp {\mbox{\boldmath $\partial$}}
\def \rb {\rm b}
\def \rg {\rm g}

\title{Field Theory of Compact Polymers on the Square Lattice}

\author{Jesper Lykke Jacobsen$^{1,2,}$%
        \footnote{E-mail: lykke@dfi.aau.dk ; lykke@thphys.ox.ac.uk}
        and Jan\'{e} Kondev$^{3,4,}$%
        \footnote{E-mail: janek@ias.edu}}

\address{$^1$Somerville College and Department of Theoretical Physics,
             University of Oxford, 1 Keble Road, Oxford OX1 3NP, U.K. \\
         $^2$Institute of Physics and Astronomy, University of Aarhus,
             Ny Munkegade, DK-8000 Aarhus C, Denmark. \\
         $^3$Institute for Advanced Study, Olden Lane, Princeton, NJ 08540. \\
         $^4$Department of Physics, Princeton University, Princeton, NJ 08540.}

\date{\today}

\maketitle

\begin{abstract}
Exact results for conformational statistics  of compact polymers 
are derived from  the two-flavour fully packed loop model on the 
square lattice.
This loop model exhibits a {\em two-dimensional manifold} of critical 
fixed points each one characterised by an infinite set of geometrical 
scaling dimensions. We calculate these dimensions {\em exactly}   
by mapping the loop model to an interface model whose scaling limit 
is described by a  Liouville field theory.  
The formulae for the central charge and the first few scaling dimensions 
are compared to numerical transfer matrix results and excellent agreement 
is found. Compact polymers are identified with a particular 
point in the phase diagram of the loop model,    
and the non-mean field value of the conformational exponent 
$\gamma = 117/112$ is calculated for the first time. 
Interacting compact polymers are described by a {\em line} of fixed 
points along which $\gamma$ varies continuously. 
\end{abstract}


\begin{multicols}{2}

\section{Introduction}

Lattice models of loops have emerged as an important paradigm in  
two-dimensional critical phenomena. They allow for a
determination of the scaling properties of different types of random walks 
which are  used to model conformations of different phases of 
polymers\cite{degennes_book}.
For instance, the solution of the O($n$)  
loop model has lead to  exact results for conformational  
exponents of swollen and dense polymers \cite{nien_On}, 
as well as polymers at the theta point \cite{saleur_theta}. The theta point 
is the tricritical point which governs the transition between the swollen and 
the collapsed phase of polymers in solution \cite{degennes_book}. 
Examples of conformational  exponents are $\gamma$, which 
describes the scaling of the number of polymer conformations with the 
number of monomers ${\cal N}$, and $\nu$,  for the scaling of the linear 
size of a polymer, as measured by the radius of gyration, 
with ${\cal N}$. Here we calculate for the first time the exact 
value of $\gamma$ for polymers on the square lattice, in the 
{\em compact phase}. Compact polymers completely fill the lattice and 
are of direct relevance to statistical studies of protein folding 
\cite{thirumalai,Chan-Dill89}. 

Further motivation for studying loop models comes from 
the Fortuin-Kasteleyn construction which maps many discrete spin models
(e.g., $Q$-state Potts) to random cluster models. 
Since cluster boundaries in two dimensions form loops this naturally 
leads to a loop model representation. This random geometrical description 
of two-dimensional lattice models then provides a setting in which a general 
theory of their scaling limits can be 
sought. It is one of the goals of this paper to outline a specific proposal 
for such a theory in the form of an effective field theory of  fluctuating 
loops. This field theory is constructed following the Coulomb gas recipe
\cite{Nienhuis} with some important new ingredients added \cite{jk_prl}. 
It describes the fluctuations of a random surface for which the loops 
are contour lines. 
 
Scaling limits of many (but not all) two-dimensional lattice models 
are described by conformally invariant field theories 
\cite{bel_pol,saleur_book}. 
This observation has lead to exact results for critical 
exponents and other  universal quantities,  and to a classification of 
critical points based on their symmetry properties with respect to the
group of conformal transformations.  
An obvious question which is often difficult to answer is: given a
particular lattice model how does one {\em construct} the conformal field 
theory of its scaling limit? Loop models provide examples for which 
the scaling limit can be constructed in a {\em physically} transparent way. 
This is accomplished  by mapping a loop model to an interface 
model, where the loops are simply equal-height  contours. 
An explicit  coarse graining procedure is then implemented for the 
height model, and it leads to a well known conformal 
field theory -- the {\em Liouville field theory}.

Interesting examples of loop models are also provided by one-dimensional 
quantum  models, spin chains in particular,  where loops appear as  
world lines of the spin. 
This mapping of spins to loops  has recently been used to formulate 
very efficient  numerical 
schemes for simulating spin chains and ladders.  These {\em loop algorithms}
allow one to simulate much bigger system sizes and lower temperatures 
than by using more traditional algorithms with local updates 
\cite{evertz_review}. 
The loop representation  of quantum spin chains also gives an illuminating
stochastic-geometrical view of their quantum fluctuations \cite{aizenman}.
For example, the spin-spin correlation function is related to the probability 
that two points on the space-time lattice belong to the same loop.  
This insight might lead to a {\em practical} theory 
of  plateau transitions in the Integer Quantum Hall Effect, i.~e., one that 
would allow for  a calculation of the correlation length exponent and other 
universal quantities which have been measured in experiments. Namely, the 
Chalker-Coddington network  model \cite{chalker},  which is believed to 
be in the same universality class as the plateau transitions,  
was recently mapped to an $SU(n\to 0)$ quantum spin chain \cite{dhlee}. 
It remains to be seen if this spin chain has a tractable loop-model 
representation. 

In the bigger picture,    
loop  models are of interest as simple examples  where the fundamental 
constituents are non-local, extended  objects as opposed to point-like objects 
such as particles and spins. Fluctuating geometries of this sort are used to  
model flux lines in superconductors, domain walls in magnets, and
crystalline interfaces, to name a few experimentally relevant systems.   

The extended nature of loops
turns out to have profound consequences when one attempts to write 
down an effective  continuum description of these models, say, following 
Landau's dictum  of expanding the free energy (Euclidean action) 
in powers of the order parameter and its derivatives. Namely, terms which are 
geometrical in origin and non-perturbative in nature,  and hence 
cannot be inferred from symmetry arguments alone,  appear in the action. 
On the other hand, exactly {\em because} these geometrical terms are present
the values of the {\em effective} coupling 
constants  of the field theory  are completely determined, 
a rather remarkable occurrence. 

Usually in an effective description provided by a field  theory, 
coupling  constants are phenomenological parameters  fixed by auxiliary 
information about observable quantities, such as the response functions or 
the related correlation functions.   
The Coulomb gas approach to two-dimensional critical phenomena is an example
of an  effective theory wherein the electromagnetic coupling constant
(i.~e., the ``magnitude of the unit charge'') is determined 
from an exact solution of the model; typically it suffices to 
calculate the exact value of a single critical exponent.
Our construction of an 
effective field theory of loop models closely parallels the Coulomb gas method 
with the important difference that the coupling constants are determined 
without recourse to any exact information about the model. For the model
at hand no such information is available anyway, and  moreover  there 
are indications that the model is not exactly solvable \cite{Batchelor96}. 
On one level 
our theory can be viewed as a trick that allows one to calculate critical 
exponents in two-dimensional loop models without doing the ``hard work'' 
of exactly solving the model. On a deeper level it shows that lattice
models of loops lead to continuum theories that are {\em geometrical} in 
nature, i.e., devoid of any couplings that depend on the microscopic details. 

\begin{figure}
 \centering\epsfig{file=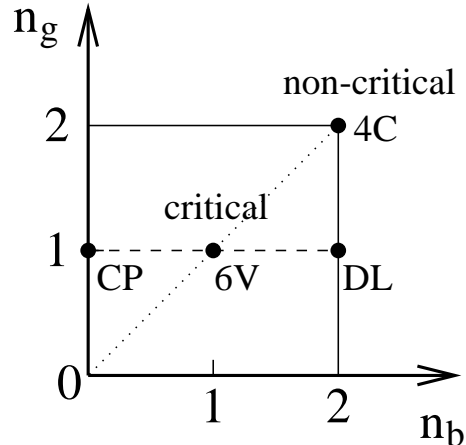,width=0.7\linewidth,angle=270}
 \begin{minipage}[t]{8.66cm}
 \caption{Phase diagram of the two-flavour fully packed loop model on the 
   square lattice. The loop model is critical for loop fugacities
   $0 \leq n_{\rb}, n_{\rg}\leq 2$.
   Particular points in the critical phase map to previously 
   studied models: 6V -- equal weighted six-vertex model [16],
   DL -- dimer loop model [15],
   4C -- four-colouring model [28].
   The dashed line is the fully packed loop model studied numerically in 
   Ref.~[14];
   the point CP along this line corresponds to the problem of 
   compact polymers. Finally, the dotted line is the loop model for which an  
   effective Liouville field theory was constructed in Ref.~[7].}  
 \label{Fig:phasediag}
 \end{minipage}
\end{figure}

Here we study in detail 
the two-flavour fully packed loop (FPL${}^2$) model on the square lattice.
This  is  
a statistical  model which describes two flavours of
loops that occupy the bonds of the square lattice, subject to certain close
packing constraints to which we shall return shortly.
The phase diagram of this model 
is described by two variables, $n_{\rm b}$ and $n_{\rm g}$, which are
the loop fugacities of the two flavours; see \Fig{Fig:phasediag}. 
The phase diagram of the FPL${}^2$ model has three
important  features that we wish to emphasize from the outset:

\noindent {\em i}) 
For loop fugacities that fall into the region 
$0\le n_{\rb}, n_{\rg} \le 2$ of the 
phase diagram  the model is critical, 
i.e., it exhibits a power-law distribution of loop sizes.  The novel 
feature is that every point in the critical region defines 
a {\em different} universality class characterised by an infinite 
set of geometrical critical exponents. All previously studied loop 
models (e.g., $Q$-state Potts, O($n$) models) exhibit a {\em line} of 
fixed points.

\noindent {\em ii})
The effective field theory of the \F2 model in the critical region 
describes a fluctuating  two-dimensional interface in 
five dimensions, which is 
characterised by {\em three} elastic constants. We calculate these three  
couplings exactly 
as a function of the two loop fugacities. It is important to note that all
previously solved loop models are characterised by a single elastic constant.
        
\noindent {\em iii}) 
{}From the field theory of the FPL${}^2$  model we calculate 
for the first time   {\em exact}  
results for the conformational exponents of compact polymers on the 
square lattice. 
Furthermore, a particular line of fixed points in the
phase diagram of the \F2 model can be identified with {\em interacting} 
compact polymers ($n_{\rb}=0, n_{\rg}\leq 2$). 
We find that along this line the  
exponent $\gamma$ changes continuously, whilst $\nu$ stays constant. 

The organisation of the paper is as follows. In Sec.~\ref{sec_com} 
we review the scaling theory of compact polymers which provides 
our main motivation for introducing the two-flavour fully packed loop
model on the square lattice in Sec.~\ref{sec_FPL}. The rest of the paper
is devoted to the study of  this model using field theoretical techniques and 
numerical transfer matrix calculations. 
 
The FPL${}^2$ model is mapped to an interface model in Sec.~\ref{sec_height}. 
For the interface model we  
construct the  scaling limit in terms of a Liouville 
field theory, in Sec.~\ref{sec_LFT}. 
In Secs.~\ref{sec_CC} and~\ref{sec_exp}
we make use of the field theory to calculate the central charge and 
the infinite set of 
geometrical exponents associated with loops, in the critical region 
of the loop model. A short description of the non-critical region based
on the field theory is given next in Sec.~\ref{sec_term}. 

Following the field theoretical treatment of the \F2 model,  
in Secs.~\ref{sec_TM} and~\ref{sec_num}  we 
describe the construction of transfer matrices for different boundary 
conditions. They are used to determine the     
central charge, the first  few geometrical exponents,
and the residual entropy; 
the numerical results are
in excellent agreement with the theoretical predictions. 
Finally, in Sec.~\ref{sec_dis}, 
we present some general observations  regarding compact polymers and 
the Coulomb gas description of conformal field theories. We also comment 
on the dimer-loop model \cite{raghavan} 
and the three-state Potts antiferromagnet \cite{widom},  in 
light of our solution of the fully packed loop model on the 
square lattice. The appendices are reserved for the  
calculation of  scaling dimensions of operators in the Liouville
field theory and the enumeration of connectivities which are used for
constructing the transfer matrices.

\section{Compact polymers}
\label{sec_com}

\begin{figure}
 \centering\epsfig{file=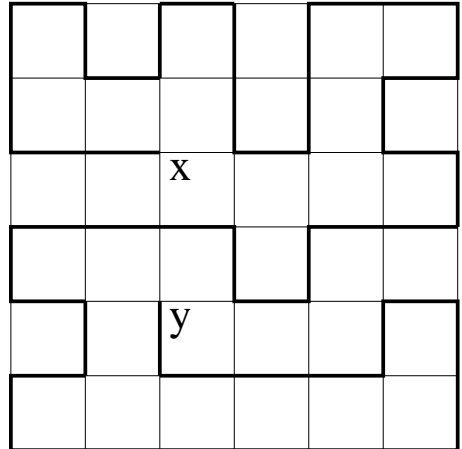,width=0.7\linewidth,angle=270}
 \begin{minipage}[t]{8.66cm}
 \caption{Compact polymer on the square lattice; ${\bf x}$ and ${\bf y}$ are 
   the positions of the chain ends.}
 \label{Fig:CP}
 \end{minipage}
\end{figure}

Compact polymers, or Hamiltonian walks, are self-avoiding 
random walks that visit {\em all} the sites of the underlying lattice; 
see \Fig{Fig:CP}. 
They have been used as simple models of polymer melts \cite{orland}
and appear in statistical studies of protein folding 
\cite{thirumalai,Chan-Dill89}. Unlike dilute and dense polymers 
whose scaling properties were calculated exactly
from  the O($n$) loop model \cite{duplantier_review}, 
compact polymers defied a similar treatment until recently. 
Numerical transfer matrix calculations \cite{nien_FPL}, a Bethe-ansatz 
solution \cite{Batch_FPL}, and 
a Coulomb gas theory \cite{jk_JPA}
of the fully packed loop  model on the {\em honeycomb} 
lattice, all conclude  that compact polymers define a new 
universality class of critical behaviour. Here we  
study compact polymers on the {\em square} lattice. We calculate 
exact scaling exponents and find them to be distinct from 
the honeycomb case.  
This was first reported in Ref.~\cite{Batchelor96} on the basis of 
numerical transfer matrix results. 

The lattice dependence of critical properties 
distinguishes the compact polymer problem from its dilute and 
dense counterparts in a crucial way. It  places them into the class of 
geometrically frustrated critical systems\footnote{Another example is the 
antiferromagnetic three-state Potts model which
has a zero-temperature critical point on the square \cite{widom} and the 
Kagom\'{e} \cite{huse_rut} lattices characterised by 
different critical exponents.}.
A physically relevant measure of
frustration for compact polymers is the number of {\em contacts} 
per monomer. Contacts are realised by monomer pairs where the 
two monomers are nearest neighbors on the lattice but are not
adjacent along the polymer chain. In lattice models of proteins
hydrophobic interactions among the amino acids occur at contacts 
\cite{thirumalai,Chan-Dill89}.  
For the square model studied here the number of contacts per monomer is
{\em two}, whilst on the honeycomb lattice it is {\em one}.

In order to study the scaling properties of compact polymers we 
focus our attention on the two most widely studied conformational   
exponents $\nu$ and $\gamma$. If $R$ is the radius of gyration of the 
polymer then
\be{nu_def}
   R \sim {\cal N}^{\nu},
\ee
where ${\cal N}$ is the number of monomers. 
Since compact polymers visit all the sites of a
lattice,  they are space-filling and  we conclude that $\nu=1/2$. This simple 
result will serve as an important check on our field theoretical 
calculations where it will be recovered.   

In order to define the conformational  exponent $\gamma$
we introduce 
$C({\cal N})$, the number of compact polymers (Hamiltonian walks) on 
a square lattice with ${\cal N}$ sites.  Since a compact polymer fills 
the lattice, boundary conditions (free, periodic, etc.) play an 
important r\^{o}le. Following Saleur and Duplantier \cite{saleur_dup_polymer},
we  define $\gamma$ in a way that is insensitive to 
the boundaries. Namely, if we introduce the quantity $C_\circ({\cal N})$, the 
number of compact-polymer {\em rings},  then we can expect
\be{gamma_def} 
\frac{C({\cal N})}{C_\circ({\cal N})} \sim {\cal N}^{\gamma},
\ee
where $\gamma$ does not depend on the choice of 
boundary conditions. Therefore,  in order to calculate $\gamma$ we need to 
solve the hard combinatorial problem of counting the number of 
open and closed compact polymers on the square lattice. Following  
de Gennes we do this by mapping the counting problem to the
calculation of a correlation function in a particular 
statistical model at the critical point. 

Consider the quantity $Z({\bf x},{\bf y}; {\cal N})$, the number 
of compact polymer conformations that start at the vertex ${\bf x}$
of the $\sqrt{{\cal N}}\times\sqrt{{\cal N}}$ square lattice, and 
end at ${\bf y}$ (see \Fig{Fig:CP}); we consider the limit
$1\ll |{\bf x}-{\bf y}|\ll\sqrt{{\cal N}}$, where ${\bf x}$ and ${\bf y}$ are
chosen far from the boundaries of the lattice. For this quantity we can 
write down the scaling form\cite{saleur_dup_polymer}:
\be{Z_scaling} 
Z({\bf x},{\bf y}; {\cal N}) = C_\circ({\cal N}) |{\bf x}-{\bf y}|^{-2x_1}
            f\left(\frac{|{\bf x}-{\bf y}|}{{\cal N}^{1/2}} \right) \ ,
\ee
where $f(u)$ is a scaling function with the property $f(u)\to {\rm const.}$
as $u\to 0$, and $x_1$ is a geometrical exponent related to $\gamma$. 
Integrating $Z({\bf x},{\bf y}; {\cal N})$ over all end-points ${\bf y}$ and 
comparing the result to \Eq{gamma_def}, the scaling relation 
\be{gama_x1} 
 \gamma = 1 - x_1
\ee
follows. 

To calculate the geometrical exponent $x_1$ we introduce the 
two-flavour fully packed loop model on the square lattice. 
The fact that we need {\em two} loop flavours follows from the simple 
observation that the bonds not covered by the compact polymer 
also form loops whose number is {\em unconstrained}. 
For the loop model we then construct an effective field theory in which 
$Z({\bf x},{\bf y}; {\cal N})$  becomes  a two-point 
correlation function. The asymptotics of this function can be 
calculated  exactly and we find $x_1=-5/112$, from which 
\be{gamma_exact} 
 \gamma = 117/112 = 1.0446\cdots
\ee
follows. This is to be compared to the mean-field theory value 
$\gamma_{\rm MF}=1$ \cite{kappa_MF}, which is also the 
result obtained for compact polymers on the honeycomb 
lattice \cite{Batch_FPL}. 

The conformational  exponent $\gamma$ was measured directly from 
enumerations of conformations of chains with lengths up to 30 in 
Ref.~\cite{thirumalai}, and the value  $\gamma=1.01(5)$ was reported. 
More recently, from a numerical transfer matrix study of the 
fully packed loop model  on the square lattice  the geometrical 
exponent $x_1=-0.0444(1)$ was determined \cite{Batchelor96}, 
in excellent agreement with the  exact result.

Another quantity of interest is the connective constant $\kappa$ which
determines the leading, exponential with system size scaling of the number 
of compact polymers \cite{Owczarek93}
\be{thirumalai-scaling}
  C({\cal N}) \sim \kappa^{\cal N} \kappa_{\rm s}^{{\cal N}^{(d-1)/d}}
                   {\cal N}^{\gamma-1} \ .
\ee
Here $\kappa_{\rm s}$ is the surface connective constant; it appears 
due to the space-filling nature of compact polymers.
Both the value $\kappa = 1.475(15)$ found in
Ref.~\cite{thirumalai}, and the estimate $\kappa \simeq 1.472$ obtained
from transfer matrix calculations similar to ours \cite{Schmalz84},
seem in favour of the mean-field result
$\kappa_{\rm MF} = \frac{4}{e} = 1.4715\cdots$ \cite{kappa_MF}.%
\footnote{Very recently the field theory of
  Ref.~\cite{kappa_MF} has been improved \cite{Higuchi98} yielding,
  however, unchanged values for $\gamma_{\rm MF}$ and $\kappa_{\rm MF}$.} 
In Sec.~\ref{sec:entropy} we report the very accurate numerical value
\begin{equation}
  \kappa = 1.472801(10),
\end{equation}
which shows that the connective constant for  compact polymers 
also deviates slightly from the mean-field result.

For the remainder of the paper we elaborate on the 
calculation of $\gamma$ for compact polymers, 
in the process unveiling an extremely rich phase
diagram of the associated loop model. As remarked earlier, it contains a 
two-dimensional region of fixed points,   which we 
characterise in detail by calculating the central charge and the  
geometrical exponents associated with loops  for each point on the 
critical manifold.

\section{Two-flavour loop  model}
\label{sec_FPL}

The two-flavour fully packed loop model on the square lattice 
was introduced in Ref.~\cite{jk_prb} as the loop representation of 
the four-colouring model~\cite{read}. It is the natural generalisation 
of the fully packed loop model on the honeycomb lattice, which
is the loop representation of the three-colouring model\cite{jk_JPA}. 
In general, a  
$q$-colouring model on a $q$-fold coordinated lattice is given by edge
colourings of the lattice with $q$ different colours; an edge
colouring of a graph is one where no two bonds that share a common
vertex are coloured equally. The colouring model is mapped to a  
loop model by choosing $[q/2]$ colour-pairs; each pair defines strings of 
alternating colour that necessarily form loops (unless they terminate 
at the boundary). In this way we end up with a loop model with  
$[q/2]$ flavours of loops.   

\begin{figure}
  \centering\epsfig{file=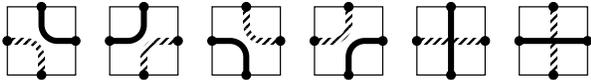,width=\linewidth,angle=0}
  \begin{minipage}[t]{8.66cm}
  \caption{The six vertex configurations of the FPL$^2$ model that are
    allowed by the fully packing constraint. Black and grey loop segments 
    are shown here as solid and hatched lines respectively. Each vertex is
    adjacent to four edges, here shown as dots, that are referred to as
    ``dangling'' if they are 
    not connected to an edge of a neighbouring vertex. Note that the two
    rightmost vertices explicitly permit the two flavours to cross.}
  \label{Fig:vertices}
  \end{minipage}
\end{figure}

To define the FPL${}^2$ model we first specify the 
allowed loop configurations ${\cal G}$. In  ${\cal G}$
every bond of the square lattice belongs to one and only one 
loop of either flavour, and loops of the same flavour are 
not allowed to cross.
Representing the two flavours by solid (black) and hatched (grey) line
segments respectively this fully packing constraint allows each vertex of
the square lattice to have one of the six appearances depicted in
Fig.~\ref{Fig:vertices}.
Each loop is assigned a fugacity depending on its flavour: 
$n_{\rm b}$ for black loops and $n_{\rm g}$ for grey loops.
The partition function of the FPL${}^2$ model is then
\be{part1}
   Z = \sum_{\cal G} n_{\rm b}^{N_{\rm b}} n_{\rm g}^{N_{\rm g}} \ . 
\ee
The fully packed loop model of Batchelor {\em et al.}\cite{Batchelor96}
is obtained by setting 
the loop fugacity of the grey loops to unity. In the limit $n_{\rm b} \to 0$ we
recover the compact polymer problem. 

If we define a restricted 
partition function of the \F2 model, to which only configurations with a 
single black loop segment propagating between points ${\bf x}$ and ${\bf y}$ 
contribute, then $Z({\bf x},{\bf y};{\cal N})$ in \Eq{Z_scaling} 
is obtained in the limit $n_{\rm b}\to 0$, $n_{\rm g}\to 1$.  
The first limit discards all configurations with black loops present,  
leaving only the black Hamiltonian walk (compact polymer) 
between ${\bf x}$ and ${\bf y}$, 
whilst the second ensures that all walks are weighted equally. We could 
also consider weighting different Hamiltonian walks  differently by 
setting $n_{\rg}\neq 1$. This situation can be 
interpreted as describing interacting 
compact polymers, and, as will be shown later, it leads to a continuously 
varying exponent $\gamma$. A similar property  
of interacting oriented polymers in the swollen  phase was suggested   
by Cardy from a field theoretical 
calculation \cite{cardy_gam}. Recent numerical studies of the 
interacting oriented self-avoiding walk by Trovato and 
Seno \cite{trovato_gam}, though, seem to be at odds with Cardy's
prediction of an exponent $\gamma$ that varies  continuously with the  
interaction strength.   

Some idea of the phase diagram of the FPL${}^2$ 
model as a function of $n_{\rm b}$ and 
$n_{\rm g}$ can be gotten by examining the extreme limits of the loop 
fugacities. Namely, for $n_{\rm b}, n_{\rm g} \to \infty$
all loops have the minimum length of four, i.e., they each surround a single
plaquette of the square lattice. There are no large loops in the 
system and the model is non-critical, or in other words, the average 
loop length is finite. 
On the other hand, in the critical phase of the loop model, which is 
the subject of this paper, in a typical configuration one finds loops of all 
sizes characterised by a power-law distribution. This leads to an 
average loop length which diverges with the system size. Such is the 
case in the other extreme limit of loop fugacities, 
$n_{\rm b}, n_{\rm g} \to 0$, when the loops cover the whole lattice.  

Other previously studied models that are particular points in the 
phase diagram of the FPL${}^2$ model are the four-colouring model,  
the dimer loop model, and the equal weighted six-vertex model; see 
\Fig{Fig:phasediag}.
For $(n_{\rm b},n_{\rm g}) = (2,2)$ the loop fugacity of each loop 
can be evenly (1+1) distributed among the two ways of colouring the 
bonds occupied by the loop with two colours in an alteranting fashion:
${\bf ABAB}\ldots$ for black loops and ${\bf CDCD}\ldots$ for grey loops. 
This is then the symmetric four-colouring model (${\bf  A,B,C,}$ and ${\bf D}$
are the colours) studied by Baxter \cite{Baxter}. 
In the dimer loop model black and white dimers are placed on the 
square lattice so that every vertex is covered by one of each \cite{raghavan}.
If we identify the dimer covered bonds with the black loops then this 
model is mapped to the $(n_{\rm b},n_{\rm g}) = (2,1)$ FPL${}^2$ model. 
And finally   $(n_{\rm b},n_{\rm g}) = (1,1)$ 
constitutes the equal-weighted six-vertex model \cite{Cardy-FSCP}, the
allowed vertices being those of Fig.~\ref{Fig:vertices}.

\section{Height representation}
\label{sec_height}

The critical phase of the FPL${}^2$ model can be  
described in terms of an effective field theory, following  the 
general procedure discussed in Ref.\cite{Nienhuis}. The idea is to 
think of loops as contours of a scalar field, which we refer to as 
the height. Depending on the loop model in question 
the height can have one or more components. If the number of components 
is $D_\perp$ then the effective field theory of the loop model describes 
a fluctuating two-dimensional interface in $D_\perp + 2$ dimensions. 

\begin{figure} 
 \centering\epsfig{file=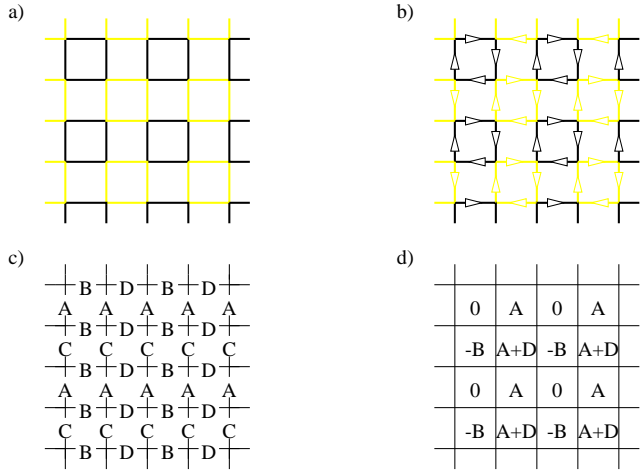,width=0.72\linewidth,angle=270}
 \begin{minipage}[t]{8.66cm}
 \caption{Mapping of the FPL${}^2$ model to an interface model. (a) $\to$ (b):
  Transform the loop configuration into an {\em oriented}  loop configuration  
  by choosing the orientation of each loop independently and randomly. 
  (b) $\to$ (c): Every 
  bond in the oriented loop configuration is in one of four states, depending
  on its flavour and direction; these four states are represented by 
  three-vectors ${\bf A, B, C}$, and  ${\bf D}$. (c) $\to$ (d): 
  The microscopic height ${\bf z}$ of 
  the interface model changes from plaquette to neighbouring 
  plaquette by ${\bf A, B, C}$,  or ${\bf D}$ depending on the state of 
  the bond between the two plaquettes. The change in ${\bf z}$ is positive 
  going clockwise around even vertices and counterclockwise around 
  odd ones.}
 \label{Fig:loop-color}
 \end{minipage}
\end{figure}

To introduce the heights we first map the loop model to an {\em oriented}
loop model, as shown in \Fig{Fig:loop-color}. 
The orientation of every loop is chosen randomly and 
independently.
Every non-oriented loop configuration is thus transformed into an 
oriented one (${\cal G}'$); the number of oriented configurations that 
correspond to the same non-oriented loop configuration is simply    
$2^{N_{\rm b} + N_{\rm g}}$. 

Next, for each loop  we redistribute its weight (fugacity), $n_{\rm b}$
or $n_{\rm g}$ depending on whether it is black or grey, between the
two possible orientations. For the black loops we do this  by
assigning to, say, the clockwise orientation the {\em phase factor} 
$\exp({\rm i}\pi e_{\rm b})$, 
and the opposite phase, $\exp(-{\rm i}\pi e_{\rm b})$,
to a counter-clockwise oriented  black loop. 
Similarly for grey loops the clockwise oriented ones are 
assigned a weight $\exp({\rm i}\pi e_{\rm g})$ whilst the counter-clockwise 
loops are weighted with $\exp(-{\rm i}\pi e_{\rm g})$. The loop fugacities are 
related to the newly introduced parameters $e_{\rm b}$ and $e_{\rm g}$ by
\bea{fug}
n_{\rm b} & = & 2 \cos(\pi e_{\rm b}) \nonumber   \\
n_{\rm g} & = & 2 \cos(\pi e_{\rm g})
\eea
since the partition function of the original (non-oriented) model, as
given by \Eq{part1}, must be recovered by independently summing
over the two possible orientations for each loop.
Note that for $0 \le n_{\rm b},n_{\rm g} \le 2$ the parameters
$e_{\rm b}$ and $e_{\rm g}$  are real, whilst for
$n_{\rm b}, n_{\rm g} > 2$ they are purely imaginary. 
As discussed in more detail in Sec.~\ref{sec_term} 
this is  the crucial property  
that leads to a critical state of the loop model 
in the former and a non-critical one in the latter case.

Now that the loops are oriented we can interpret them as contours of a
height field; the orientation is necessary as it determines the direction 
of increasing height. 
The systematic construction of the {\em microscopic heights} sets out from the 
observation that every bond of the square 
lattice is in one of four possible states: it can be  coloured   
black or grey, and oriented from an even to an odd site, 
or from odd to even. ``Even'' and ``odd'' refer here to the 
two sublattices of the bipartite square lattice;  every even site
is surrounded by four nearest neighbouring odd sites, and {\em vice versa}.

The four possible bond-states are represented by four vectors -- which  are 
the colours in the four-colouring representation -- ${\bf A},
{\bf B}, {\bf C}$ and ${\bf D}$; see \Fig{Fig:loop-color}c. 
The microscopic heights $\{{\bf z}\}$ are defined on the dual 
lattice and  the change in height when going from 
one plaquette centre to the next is given  by 
${\bf A},{\bf B}, {\bf C}$ or ${\bf D}$, depending 
on the state of the bond which is crossed; \Fig{Fig:loop-color}d. 
For the height to be uniquely 
defined the four vectors must satisfy the constraint 
${\bf A}+{\bf B}+{\bf C}+{\bf D}=0$. This means that the microscopic 
heights live in a {\em three-dimensional} vector space, which we take to be 
$\Zs^3$. In other words the oriented \F2 model maps to a model of a 
two-dimensional interface in five spatial dimensions. 

By reasons of symmetry the four vectors are chosen so as to
point from the centre to the
vertices of a regular tetrahedron.
With a suitable choice of 
coordinates they are represented by three-vectors: 
\bea{colour-vec}
 {\bf A} & = & (-1,+1,+1) \nonumber \\ 
 {\bf B} & = & (+1,+1,-1) \nonumber \\
 {\bf C} & = & (-1,-1,-1) \nonumber \\ 
 {\bf D} & = & (+1,-1,+1) \ .
\eea
This is the same normalisation as the one used in Ref.~\cite{jk_prb}. 

Mapping the loop model to on oriented loop model also allows for a 
{\em local} redistribution of the loop weights. This is important since 
it leads to a local field theory for the heights. 
As we will find out shortly, though local, this field theory  
is somewhat unconventional due to the non-local, extended nature of the 
fundamental microscopic objects it purports to describe.   

To redistribute the phase factors associated with oriented loops we
assign a phase $\exp(-i\pi e_{\rm b}/4)$ to a vertex of the square lattice
if a black loop makes a left turn at that vertex, the opposite phase 
$\exp(+i\pi e_{\rm b}/4)$ if it makes a right turn, and the weight $1$ if it 
continues straight. The total vertex weight $\lambda({\bf x})$ 
is a product of the phase factor originating from the black loop and an 
equivalent one from the grey loop passing through the same 
vertex ${\bf x}$. The partition function of the FPL${}^2$ model,
\Eq{part1}, can  now be rewritten as a sum over oriented loop
configurations (i.e., colouring configurations) 
\be{part2}
 Z = \sum_{{\cal G}'} \prod_{\bf x} \lambda({\bf x}) \ .
\ee
Once the height at a single point is fixed ${\cal G}'$ is in a one-to-one
correspondence with the configurations of the microscopic heights, and the 
summand in the above equation is the appropriate weight.
In the critical phase of the FPL${}^2$ model the interface described by 
\Eq{part2} is {\em rough},  and the 
field theory is constructed so as to correctly reproduce its long-wavelength
fluctuations.   

\subsection{Spectrum of electromagnetic charges} 

The mapping from oriented loop configurations, which are equivalent to 
edge colourings,  to microscopic height configurations is one to 
many. In particular, two height configurations corresponding to the 
same edge colouring can have their heights shifted with respect to each other 
by a {\em global} shift ${\bf m}\in{\cal R}$. 
The set ${\cal R}$ forms a three-dimensional Bravais lattice, i.e., it 
is closed under integral linear combinations, and its elements are the   
{\em magnetic} charges in the Coulomb gas representation of the
FPL${}^2$ model.
The lattice reciprocal to the lattice of magnetic charges, ${\cal R}^*$,  
defines the {\em electric} charges ${\bf e}\in{\cal R}^*$, with the property 
${\bf e}\cdot{\bf m} = 2\pi m, m\in\Zs$. 

The construction of the lattice ${\cal R}$ for the FPL${}^2$ model 
follows the usual prescription 
for height models, and has been carried out in detail in Ref.\cite{jk_prb}. 
For the sake of completeness we outline this construction below. 
  
It is convenient to first identify the {\em flat} 
states (also referred to as the {\em ideal} states), i.e.,  
those colouring states which minimise the variance of the 
microscopic height ${\bf z}$. 
{}From the height mapping described above it follows that these 
states have all of their 
plaquettes coloured with two colours only; 
an example is shown in \Fig{Fig:loop-color}c. 
This leads to a colouring 
state that is periodic,  with the same $2\times 2$ colouring pattern 
repeated throughout the lattice.   There are twenty four flat/ideal 
states for the colouring representation of the FPL${}^2$ model, corresponding 
to the number of permutations of four different colours. Namely,  an 
ideal state is completely specified by listing the colours of the 
bonds around a single site (say the origin),   
starting from the left horizontal bond and proceeding clockwise.   
To each flat  state we  assign a {\em coarse grained height}
${\bf h} = \langle {\bf z} \rangle$, 
which is the average microscopic height over a $2\times 2$ unit cell
of the colouring. 

The flat states form a three dimensional graph, which we refer to as the 
ideal state graph, ${\cal I}$. Namely,  
starting from any ideal state four other ideal states can be reached  
by exchanging a pair of  colours that form  a plaquette. 
For example,  by exchanging the colours ${\bf A}$  and ${\bf B}$ 
in \Fig{Fig:loop-color}c all the ${\bf A}{\bf B}{\bf A}{\bf B}$ 
plaquettes are turned  into ${\bf B}{\bf A}{\bf B}{\bf A}$  
plaquettes to give a new ideal state.
Under these  plaquette flips only the microscopic heights at the centres of
the  affected plaquettes are changed. In this way the ideal states form a 
four-fold coordinated graph in height space, where each vertex is 
indexed by a colour permutation, and its position in $\Rs^3$  is 
given by the coarse grained height ${\bf h}$. Bonds are associated with 
transpositions of two colours; they lie along the direction defined by 
the difference of the two colour vectors, and have a length of $\sqrt{2}/2$ 
if the normalisation in \Eq{colour-vec} is chosen.

\begin{figure} 
 \centering\epsfig{file=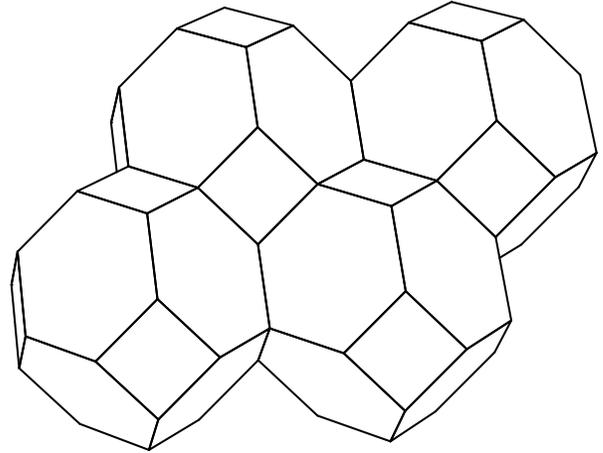,width=\linewidth,angle=0}
 \begin{minipage}[t]{8.66cm}
 \caption{The ideal state graph of the \F2 model in the four-colouring 
  representation.} 
 \label{Fig:polyhedra}
 \end{minipage}
\end{figure}

The ideal state graph  is a tiling of 
$\Rs^3$ with truncated octahedra; this regular polyhedron  
is  better known as the 
Wigner-Seitz cell of a body-centred cubic (bcc) 
lattice (see \Fig{Fig:polyhedra}). 
A single truncated octahedron  in ${\cal I}$ 
has twenty four vertices corresponding 
to the twenty four different ideal states. The set of vertices in 
${\cal I}$ representing 
the same ideal state form the {\em repeat lattice} ${\cal R}$, which 
is face-centred cubic (fcc) with a conventional cubic cell of side 4.

To obtain the continuum description of the FPL${}^2$ model we coarse 
grain the microscopic height over domains of ideal states. This gives 
rise to the coarse grained height ${\bf h}$ which we can 
consider to be {\em compactified} on $\Rs^3/{\cal R}$. The phase space
of the height is not simply connected, thus allowing for topological 
defects (vortices) with topological charges that take their values in 
${\cal R}$\cite{chaikin-lubensky}.  
These defects are associated with  {\em magnetic} charges in the 
Coulomb gas representation of the FPL${}^2$ model. {\em Electric} 
charges on the other hand are associated with 
vertex operators $\exp({\rm i}{\bf e}\cdot{\bf h})$. If we take the 
height to live in $\Rs^3/{\cal R}$ then vertex operators are well defined 
only for values of the electric charge ${\bf e}\in{\cal R}^*$. 
${\cal R}^*$ is the lattice {\em reciprocal} to the lattice of 
magnetic charges ${\cal R}$, 
and it is a  body-centred cubic (bcc) lattice 
with a conventional cubic cell of side $\pi$.

\section{Construction of the field theory}
\label{sec_LFT}

An effective field theory of the FPL${}^2$ model should describe large scale 
properties of loops. The kind of questions we expect it to answer 
are ones that do not refer to  the microscopic details of the lattice
model. For  example, from the effective field theory we will   
calculate the asymptotics of the probability that two 
points lie on the same loop,  when the separation between the 
points is large  compared to the lattice spacing. {}From this and related 
quantities the conformational exponents of compact polymers can be 
extracted.

The field theory of the \F2 model 
is defined by the Euclidean action for the coarse grained 
height ${\bf h}$.  
Consider a typical configuration of the oriented FPL${}^2$ model which is 
equivalent to the colouring model. It consists of domains of ideal states. 
To each ideal state domain we assign a coarse grained height, defined earlier 
as the average microscopic height over the domain. In the continuum limit
we assume that this height is a smoothly varying function of the 
basal plane coordinates $(x^1, x^2)$. The partition function 
that takes into account only the large scale 
fluctuations of the height can be written as a functional  integral,
\be{path}
  Z_> = \int \! {\cal D}{\bf h} \ \exp(-S[{\bf h}]),
\ee
where $S$ is the Euclidean action of a Liouville field theory 
with imaginary couplings \cite{jk_prl}.
The Liouville action contains three terms, 
\be{action_tot}
S = S_{\rm E} + S_{\rm B} + S_{\rm L} \ .
\ee
Each one has  a concrete geometrical interpretation in the 
FPL${}^2$ model, which we describe next.

\subsection{Elastic term}
\label{sub_ET}

\begin{figure} 
 \centering\epsfig{file=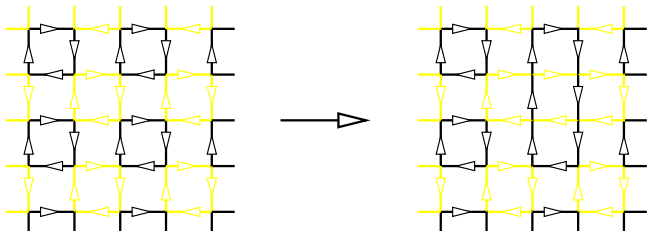,width=0.36\linewidth,angle=270}
 \begin{minipage}[t]{8.66cm}
 \caption{A loop flip changes one oriented loop configuration into another. 
  Here the bond states ${\bf C}$ and ${\bf B}$ are exchanged along a single 
  ${\bf BC}$ plaquette (cfr.~\Fig{Fig:loop-color}c).}
 \label{Fig:loop_flip}
 \end{minipage}
\end{figure}

The first term in the effective action for the FPL${}^2$ model describes the 
elastic fluctuations  of the interface. It gives less weight to 
configurations that deviate from the  flat states, 
by penalising finite  gradients of the height. This term is entropic in origin.
Namely, in order to change the colour of a particular 
bond in the four-colouring representation of the loop model, 
say ${\bf C}\rightarrow{\bf B}$, all the ${\bf C}$'s and ${\bf B}$'s
have to be interchanged along the ${\bf CB}$ loop which contains the 
chosen bond. This transformation we call a {\em loop flip}; see  
\Fig{Fig:loop_flip}. 
The ideal states {\em maximise} the number of loops of 
alternating colour and consequently they have the largest entropy of 
loop flips.  

In its most general form the elastic term in the effective 
action can be written as a gradient expansion, 
\be{action_el}
S_{\rm E} = \frac{1}{2} \int \! d^2{\bf x}   K^{ij}_{{\alpha}{\beta}}  
            \partial_i h^{\alpha} \partial_j h^{\beta} \ ,
\ee
where higher powers of the height gradients and higher derivatives of the 
height are less relevant at 
large scales. The stiffness tensor $K^{ij}_{{\alpha}{\beta}}$ nominally has
36 components; the indices $i,j=1,2$ are for the basal plane coordinates, 
whilst  $\alpha,\beta=1,2,3$  label the three components of the
height. Summation over repeated indices is assumed throughout. 

The number of independent non-zero components of the stiffness tensor 
(i.e., elastic constants) is 
actually only {\em three}, once all the symmetries of the FPL${}^2$ model are 
taken into account. The relevant symmetry transformations, that is  the 
ones that become the symmetries of the effective action,   are the ones 
that leave the weights of oriented loop configurations unchanged.  
First, there are the lattice symmetries, 
translations and rotations, which cut the number of independent 
elastic constants down to six. The terms that are allowed in $S_{\rm E}$ are 
scalars under rotations in the basal plane $\{(x^1, x^2)\}$, and they are 
necessarily of the form $\bp h^{\alpha}\cdot \bp h^{\beta}$, where 
$\bp=(\partial_1, \partial_2)$ is the usual gradient.
Second,  the FPL${}^2$ model possesses colour symmetries,  
\be{sym1} 
 {\bf A} \leftrightarrow {\bf B}:
   e_{\rm b} \leftrightarrow -e_{\rm b} \ {\rm and} \ 
   z_1 \leftrightarrow z_3
\ee
and 
\be{sym2}
 {\bf C} \leftrightarrow {\bf D}:
   e_{\rm g} \leftrightarrow -e_{\rm g} \ {\rm and} \ 
   z_1 \leftrightarrow -z_3 \ ,
\ee
which interchange the colours and at 
the same time transform the microscopic heights. 
Taking into account the colour symmetries the elastic 
contribution to the action 
takes on the form:
\bea{action_el_sym}
S_{\rm E} = \frac{1}{2} \int \! d^2{\bf x} \ \left \lbrace
      K_{11}[(\bp h^1)^2 + (\bp h^3)^2] \right. &+& \\ 
      2 K_{13}(\bp h^1 \cdot \bp h^3) &+& \left. K_{22} (\bp h^2)^2
      \right \rbrace \nonumber.
\eea
Furthermore, by introducing a change of coordinates in height space, 
\be{quadr}
H^1 = \frac{1}{2}(h^1 - h^3) \ , \ \ 
H^2 = h^2 \ , \ \ 
H^3 = \frac{1}{2} (h^1 + h^3) 
\ee
$S_{\rm E}$ becomes diagonal,     
\be{action_el_sym2}
S_{\rm E} = \frac{1}{2} \int \! d^2{\bf x} \ g_{\alpha} (\bp H^{\alpha})^2 \ .
\ee
The three coupling constants $g_{\alpha}$ ($\alpha = 1,2,3$) are
linearly related to the three elastic constants, 
\be{coup}
  g_1  =  2 (K_{11} -  K_{13}),  \ \
  g_2  =  K_{22},   \ \
  g_3  =  2 (K_{11} + K_{13}).
\ee 

The appearance  of {\em three} elastic constants is rather 
intriguing from the viewpoint of loop models that have been solved 
previously. The $Q$-state Potts, the O($n$), and the honeycomb FPL models
are all characterised by a {\em single} coupling constant, which has been  
determined case by case from their exact solutions.   
Below we will show that all  three couplings in 
\Eq{action_el_sym2} can be 
calculated exactly from the {\em loop ansatz} introduced in 
Ref.~\cite{jk_prl}.\footnote{The coupling constant $g$ for all the 
loop models known to date can be calculated using this method, therefore 
dispensing with the need for an exact solution.} 
The ansatz states that the operator which enforces the 
complex  weights assigned to oriented loops is 
{\em marginal} in the renormalisation group sense. This property of 
the field theory is intimately related to the random geometry of loops;
we elaborate on this important point in Section~\ref{loop_ansatz}.

\subsection{Boundary term}

The mapping of the loop model to an oriented loop model with local complex 
weights $\lambda({\bf x})$ (\Eq{part2}) 
fails for loops that experience the boundary. 
For example, if we define the FPL${}^2$ model on a 
cylinder then loops that wind around the cylinder will not be weighted
properly. The winding loop has an equal number of left and right turns and 
hence it  will be assigned a weight one. Summing over the two orientations
gives a weight two, and not the correct $n_{\rm b}$ or $n_{\rm g}$,
depending on the flavour. To correctly weight these loops one introduces a 
boundary term into the effective action,  
\be{action_bc}
 S_{\rm B} = \frac{{\rm i}}{4 \pi} \int \! d^2{\bf x} \ ({\bf e}_0 
             \cdot {\bf h}) {\cal R} \ ;
\ee
${\cal R}$ is the scalar curvature and ${\bf e}_0$ is the {\em 
background} electric charge, which is to be determined. Since we are 
only concerned with the situation where the lattice on which the FPL${}^2$ 
model is defined is flat, the scalar curvature vanishes everywhere 
except at the boundary.  

To determine ${\bf e}_0$ we consider the FPL${}^2$ model on the 
cylinder. The scalar curvature of the cylinder is proportional to the
difference of two delta functions situated at the two far ends of the
cylinder: 
${\cal R} = 4 \pi \left[\delta(+\infty) - \delta(-\infty) \right]$.   
Therefore $S_{\rm B}$  has the effect of placing vertex operators
$\exp(\pm{\rm i} {\bf e}_0 \cdot {\bf h})$ at $x^2= \pm \infty$;
here $x^2$ is the coordinate along the length of the cylinder. 
These  vertex operators assign an additional  weight $\exp({\rm i}
{\bf e}_0 \cdot ({\bf h}(+\infty) - {\bf h}(-\infty))$ to oriented loop 
configurations on the cylinder. Now, in order for  
${\bf h}(+\infty) - {\bf h}(-\infty)$  to be  non-zero there must be at 
least a single winding loop present. If this winding loop is black, then the 
height difference is ${\bf A}$ or ${\bf B}$ depending on its orientation; 
similarly if the loop is grey the height difference is ${\bf C}$ or ${\bf D}$.
Furthermore if the background charge is chosen so as to satisfy  
\bea{bc-eqs} 
 {\bf e}_0 \cdot {\bf A} =   \pi e_{\rm b} \ \ \ \ &
 {\bf e}_0 \cdot {\bf B} = - \pi e_{\rm b} \nonumber \\ 
 {\bf e}_0 \cdot {\bf C} =   \pi e_{\rm g} \ \ \ \ &
 {\bf e}_0 \cdot {\bf D} = - \pi e_{\rm g} 
\eea
then the winding loops will be assigned their proper weights.   
This is again seen by summing over the two possible orientations of
the winding loop.
In the normalisation chosen for the colour vectors, \Eq{colour-vec}, 
the unique solution of the system of linear equations in \Eq{bc-eqs} is 
\be{bc} 
 {\bf e}_0 = - \frac{\pi}{2} (e_{\rm g} + e_{\rm b}, 0, e_{\rm g} - e_{\rm b}).
\ee
This calculation of the {\em vector} background charge generalises the scalar 
case studied previously \cite{Nienhuis}.

\subsection{Liouville potential}

The elastic term and the boundary term make up  
the usual Coulomb gas approach to  two-dimensional critical phenomena. 
Recently we have argued that this description is incomplete and that an
extra  term $S_{\rm L}$ must be added to the effective action.
To see this consider a large loop in the bulk, one that does
not experience the boundary. Without the extra term this loop would be
weighted exclusively by the bulk term $S_{\rm E}$. There are two
problems with this: $S_{\rm E}$ is real whilst an oriented loop should be
weighted by a complex phase, and, $S_{\rm E}$ 
does not distinguish between the two orientations of a loop which are
assigned different weights. 
We conclude that an extra {\em bulk} term is necessary!

The most general form of a bulk term is  
\be{action_lp}
S_{\rm L} = \int \! d^2{\bf x} \ w[{\bf h}({\bf x})] \ ,
\ee
where $\exp(-w[{\bf h}({\bf x})])$ is the scaling limit of 
$\lambda({\bf x})$ in \Eq{part2}. In this sense $S_{\rm L}$ 
is energetic in origin, as opposed to $S_{\rm E}$, which we argued in 
Sec.~\ref{sub_ET} accounts for the entropy of edge colourings.  

Microscopically, the vertex weight $\lambda$ can be written in 
terms of the colours of the bonds around  the particular vertex as 
$\lambda=\exp(-w)$ where
\bea{weight_op}
w({\bf B}, {\bf C}, {\bf A}, {\bf D}) & = &  0, \nonumber \\
w({\bf B}, {\bf D}, {\bf A}, {\bf C}) & = &  0, \nonumber \\
w({\bf A}, {\bf B}, {\bf C}, {\bf D}) & = & \mp {\rm i} \frac{\pi}{4}
                                           (e_{\rm g}+e_{\rm b}), \nonumber \\
w({\bf B}, {\bf A}, {\bf C}, {\bf D}) & = & \mp {\rm i} \frac{\pi}{4}
                                           (e_{\rm g}-e_{\rm b}), \nonumber \\
w({\bf A}, {\bf B}, {\bf D}, {\bf C}) & = & \mp {\rm i} \frac{\pi}{4}
                                           (e_{\rm b}-e_{\rm g}), \nonumber \\
w({\bf B}, {\bf A}, {\bf D}, {\bf C}) & = & \mp {\rm i} \frac{\pi}{4}
                                           (-e_{\rm b}-e_{\rm g}) \ ;
\eea
the top sign is for even vertices whilst the bottom sign applies
to odd vertices of the square lattice. Here we adopt the notation
$(\mathbf{\sigma}_1,\mathbf{\sigma}_2,\mathbf{\sigma}_3,\mathbf{\sigma}_4)$
for the ordering of the colours around a vertex by listing the colours 
clockwise from the leftmost bond. The operator $w$ is completely
specified by the values it takes on the six edge colourings listed above 
since it does not change under cyclic permutations of its arguments.

By explicitly going through the six colour configurations listed above
it is easily checked that
\be{weight_general}
 w({\bf x}) = \frac{{\rm i}}{16} \ {\bf e}_0 \cdot {\bf Q}({\bf x}),
\ee
where the {\rm cross-staggered} operator\cite{jk_prb} is defined by
\be{cross_staggered}
  {\bf Q}({\bf x}) = \pm [\mathbf{\sigma}_1({\bf x}) - 
              \mathbf{\sigma}_3({\bf x})]
              \times [\mathbf{\sigma}_2({\bf x}) - \mathbf{\sigma}_4({\bf x})].
\ee
Since ${\bf Q}({\bf x})$ is manifestly invariant under $90^\circ$ rotations
of the colours around ${\bf x}$, \Eq{weight_general} is seen to hold true
for any distribution of the colours around a given vertex.

In order to find the coarse grained version of $w({\bf x})$ we 
express it as a function of the height field ${\bf h}({\bf x})$. 
First note that the microscopic operator $w({\bf x})$ 
is {\em uniform} in each of the ideal states 
of the four colouring model. As such it defines a function on the
ideal state graph $w({\bf h})$, where ${\bf h}\in{\cal I}$ is the 
coarse grained height. Furthermore, it is a periodic
function of ${\bf h}$ and it can therefore be written as a Fourier sum:
\be{F_sum}
w({\bf h}) = \sum_{{\bf e}\in{\cal R}^*_{w}} 
         \tilde{w}_{\bf e} \exp({\rm i} {\bf e} \cdot {\bf h}) \ . 
\ee
The electric charges appearing in the sum take their values in the 
sub-lattice ${\cal R}^*_w \subset {\cal R}^*$, which is the lattice 
reciprocal to the lattice of {\em periods} of $w({\bf h})$. In the continuum 
limit the coarse-grained height ${\bf h}$ is promoted into the height
field ${\bf h}({\bf x})$, and the scaling limit of the operator $w$ is 
obtained by replacing ${\bf h}$ by ${\bf h}({\bf x})$ in \Eq{F_sum}.
Therefore $w[{\bf h}({\bf x})]$ is a sum of vertex operators,  
\be{F_sum2}
 w[{\bf h}({\bf x})] = 
   \sum_{{\bf e}\in{\cal R}^*_w} \tilde{w}_{\bf e} \exp({\rm i}
   {\bf e} \cdot {\bf h}({\bf x})) \ , 
\ee 
of which  only the most relevant one(s) are kept in the effective action. Since
the relevance of an operator is determined by its scaling dimension we 
turn to this calculation next.

\subsubsection{Dimensions of charge operators}

In the Coulomb gas formalism operators are associated with
either electric or magnetic charges.
Electric operators are vertex operators $\exp({\rm i}{\bf e}\cdot{\bf h})$
and they appear as the  scaling limits of microscopic operators 
in the FPL${}^2$ model that can be
expressed as local functions of the colours; the loop-weight 
operator is one example.
 
Magnetic operators on the other hand  
cannot be expressed as local functions of the 
height but can be thought of as  a 
constraint on the height field that generates a topological defect of 
strength ${\bf m}$. If  ${\bf x}$ is the position of 
the defect core then  the  
net height increase around any loop that encloses ${\bf x}$  
is ${\bf m}$ (assuming no other defects are encircled). 
Geometrical  exponents for loops in the 
FPL${}^2$ model are given by  dimensions of electric and 
magnetic operators in the associated Coulomb gas.   

For an operator that has  
total electromagnetic charge $({\bf e}, {\bf m})$, where  
${\bf e}=(e_1,e_2,e_3)$ and ${\bf m}=(m^1,m^2,m^3)$, the scaling dimension 
is the sum of the electric and magnetic dimensions,\footnote{The 
derivation of \Eq{em_dim} is an exercise in Gaussian integration 
and is reviewed  in Appendix \ref{app:Gauss}.}
\be{em_dim}
 2 x({\bf e}, {\bf m}) = \frac{1}{2 \pi} \left[ \frac{1}{g_\alpha} 
       E_\alpha(E_\alpha - 2 E_{0\alpha}) + g_\alpha (M^\alpha)^2 \right] \ , 
\ee
where 
\be{transf}
E_1 = e_1 - e_3 \ , \ \ 
E_2 = e_2 \ , \ \ 
E_3 = e_1 + e_3
\ee 
and 
\be{quadr1}
M^1 = \frac{1}{2}(m^1 - m^3) \ , \ \ 
M^2 = m^2 \ , \ \ 
M^3 = \frac{1}{2} (m^1 + m^3) 
\ee
are the electric and magnetic charge vectors in the basis
in which the elastic term in the action is diagonal.
Since the magnetic charges are given by 
height differences they 
must transform according to \Eq{quadr}, whilst the electric
charges transform in a dual fashion (cfr.~their appearance in the
vertex operators).

\subsubsection{Loop ansatz}
\label{loop_ansatz}

With the dimension formula  
in hand,  we can settle the issue of the most 
relevant operators appearing in the Fourier expansion of $w({\bf h})$; 
see \Eq{F_sum2}. There are twelve vertex operators to choose from corresponding
to the twelve (110)-type vectors in the bcc lattice ${\cal R}^*$; these are 
the shortest vectors in the lattice ${\cal R}^*_w$. To find
which of these electric charges minimise  $x({\bf e},0)$ (\Eq{em_dim})
it is convenient to
first consider  the simpler case of the FPL${}^2$ model for
$n_{\rm b} = n_{\rm g}$.

For the FPL${}^2$ model with equal fugacities for the black and grey loops  
the effective action is considerably simplified. Namely, in this case
the cyclic permutation of the colours,  
\bea{cycl-symm}
  ({\bf A},{\bf B},{\bf C},{\bf D}) & \leftrightarrow &
  ({\bf B},{\bf C},{\bf D},{\bf A}) \ :  \nonumber \\
  (z_1,z_2,z_3) & \leftrightarrow & (-z_1,z_3,-z_2)
\eea
does not change the vertex weight 
$\lambda$, and is thus an  additional symmetry of the action $S$.  
This symmetry implies 
that $K_{13} = 0$  and $K_{22} = K_{11}$  in \Eq{action_el_sym}. 
Consequently there is only one elastic constant, $K \equiv K_{11}$. 
This then 
simplifies the formula for the dimension of an electromagnetic charge,
\be{em_dim2} 
2 x({\bf e}, {\bf m}) = \frac{1}{2\pi K} {\bf e}\cdot({\bf e} - 2{\bf e}_0)
                        + \frac{K}{2\pi} {\bf m}^2 \ ,
\ee
where from \Eq{bc} it follows that   
the background charge in this case has only one non-zero component,
${\bf e}_0=-\pi(e_{\rm b},0,0)$.  Now it is a simple matter to 
check that of the twelve 
(110)-type vectors in the lattice of electric charges  
${\cal R}^*$, the four charges  
\bea{marg_ch}
{\bf e}^{(1)} & = & (-\pi, 0, +\pi), \nonumber \\
{\bf e}^{(2)} & = & (-\pi, 0, -\pi), \nonumber \\
{\bf e}^{(3)} & = & (-\pi, +\pi, 0), \nonumber \\
{\bf e}^{(4)} & = & (-\pi, -\pi, 0) 
\eea 
are degenerate in dimension and they minimise $2x({\bf e}, 0)$. 
These are therefore the  electric charges of the vertex operators that 
are kept  in the action.   

Now we turn to the {\em loop ansatz} which states that the operator
$w({\bf h})$ is exactly marginal in the renormalisation group sense. This is 
the statement that the loop weight does not renormalise at large scales. 
The geometrical meaning of this becomes obvious when one realises
that the number of loops inside a domain of size $\rho$,  whose linear size 
is {\em comparable} to $\rho$, is thermodynamically conjugate to 
the loop weight at scale $\rho$. Thus the  
loop ansatz states that the number of large loops does not grow with scale
(more precisely it is sufficient to assume that it does not grow 
faster than any power of the scale). 
The analogous statement can be proven rigorously  
for critical percolation where it is the source of hyperscaling \cite{aiz2}.

The assumption that there is of order one loop at every scale is linked 
to the variance of the height difference between two points in the 
basal plane, separated by a macroscopic distance $|{\bf x}|$. 
Namely, if we assume that when going from 
one point to the other there is of order one contour loop that is crossed at 
every {\em scale}, and further assuming that the directions of these  
contours are independent from scale to scale, it follows from the 
law of large numbers   
that the variance of the height difference grows as 
the number of contours crossed, that is as  
$\log(|{\bf x}|)$. 
This of course is nothing but the 
large $|{\bf x}|$ behaviour of
$\left<(H^\alpha({\bf x})-H^\alpha(0))^2\right>$  
calculated in the Gaussian model of \Eq{action_el_sym2}. 

The loop ansatz, or in other words the marginality hypothesis for the 
loop weight operator, simply translates into a statement about its
scaling dimension:
\be{marg_dim}
  x({\bf e}^{(i)}, 0) = 2 \ \ \ i=1,2,3,4 \ .
\ee
This, using the dimension formula \Eq{em_dim2}, leads to a formula for the 
single elastic constant $K$. 

In the general case $n_{\rm b}\neq n_{\rg}$, the scaling dimensions of 
the four electric charges identified above are
\bea{ei_dims}
x({\bf e}^{(1)}, 0)  &  = &  \pi \ \frac{1-e_{\rb}}{g_1}, \nonumber \\
x({\bf e}^{(2)}, 0)  &  =  & \pi \ \frac{1-e_{\rg}}{g_3},           \\
x({\bf e}^{(3)}, 0)= x({\bf e}^{(4)}, 0) & = & 
\frac{\pi}{4} \ \left( \frac{1-2e_{\rb}}{g_1} + \frac{1}{g_2} + 
 \frac{1-2e_{\rg}}{g_3} \right) ; \nonumber 
\eea
the last two remain degenerate in dimension. 
The dimensions of the first two charges are also equal due to the 
``duality'' transformation of the \F2 model which exchanges the two 
flavours, $n_{\rb}\leftrightarrow n_{\rg}$. This transforms the 
microscopic heights $z_2 \rightarrow -z_2$ and $z_3 \rightarrow -z_3$ 
(and similarly for the appropriate components of the height field).  
Furthermore, the elastic constants $K_{11}$ and $K_{22}$ in \Eq{action_el_sym} 
are unchanged, whilst  $K_{13} \rightarrow - K_{13}$. 
Finally, from \Eq{coup} it follows  that the 
duality transformation exchanges the couplings $g_1 \leftrightarrow g_3$
thus rendering ${\bf e}^{(1)}$ and ${\bf e}^{(2)}$ degenerate in dimension, 
as the \F2 model is self-dual.  

Unlike the case of $n_{\rm b}= n_{\rg}$, the loop ansatz in the
general case requires that at least {\em two} of 
the electric charges ${\bf e}^{(i)}$ ($i=1,2,3,4$) remain marginal, 
thus enforcing 
the non-renormalisability of the two fugacities $n_{\rb}$ and $n_{\rg}$. 
If we now further {\em assume} that these charges are unrelated by the 
``duality'' transformation described above, it follows that in fact all 
four are marginal. The three couplings are then simply calculated by 
setting the right hand sides of \Eq{ei_dims} equal to 2. We find: 
\bea{couplingS}
 g_1 & = & \frac{\pi}{2} (1 - e_{\rm b}), \nonumber \\
 g_3 & = & \frac{\pi}{2} (1 - e_{\rm g}), \\
 \frac{1}{g_2} & = & \frac{1}{g_1} + \frac{1}{g_3} \ . \nonumber
\eea

One final comment is in order.   
The relation $1/g_2= 1/g_1 + 1/g_3$ comes as somewhat of a surprise, 
as it was  
not anticipated on symmetry grounds. Of course, since a particular point in 
the critical region of the FPL${}^2$ model is determined by two parameters,  
$n_{\rm b}$ and $n_{\rm g}$,  one relation between the three couplings is to 
be expected. It is therefore an interesting open question whether a critical 
loop model can be constructed in which 
$g_2$ would be unconstrained.\footnote{This possibility was 
suggested to us by D.~Huse.} 

With the values of the couplings $g_1, g_2$, and $g_3$ in hand, as well as 
the formula for the scaling dimensions of charged operators, 
\Eq{em_dim}, we are fully equipped to  
calculate critical exponents of the FPL${}^2$ model.
In particular, in the next section we calculate  
the formulae for the central charge and
the geometrical exponents associated with loops as a function of the 
loop fugacities, $n_{\rb}$ and $n_{\rg}$, 
for the whole critical region of the model.

\section{Central Charge}
\label{sec_CC}

We now turn to the calculation  of the central charge in the critical region,
$0 \le n_{\rm b},n_{\rm g} \le 2$. Exactly at the point
$(n_{\rm b},n_{\rm g}) = (2,2)$ the background charge vanishes,
${\bf e}_0 = {\bf 0}$, and the action consists only of the elastic
term $S_{\rm E}$ given by \Eq{action_el_sym2}. Since this is then simply a
theory of three free massless bosonic fields we conclude that, in this
case, $c=3$ \cite{jk_prb}.

For a general value of the background charge this generalises to 
\cite{dots_fateev}
\be{c_charge}
c = 3 + 12 x({\bf e}_0, 0) \ .
\ee
One way to rationalise the factor of 12 is to compare  the  
coefficients of the finite-size corrections in the well-known formulae
\cite{Cardy86,Cardy83}
\bea{fss_corr}
 f_0(\infty) - f_0(L) &=& \frac{\pi c}{6L^2} + \cdots \\
 f_i(L) - f_0(L)      &=& \frac{2\pi x_i}{L^2} + \cdots \ , 
\eea
where $f_{0,i}(L)$ is the free energy density 
on a cylinder of circumference $L$, the subscript $0$ referring 
to the vacuum and $i$ to the case 
when an operator of scaling dimension $x_i$ is inserted. The physical 
meaning of \Eq{c_charge} is that the  
presence of the background charge -- $+{\bf e}_0$  and $-{\bf e}_0$
at the two ends of the 
cylinder -- lowers the free energy and with it the central charge.

Now using the dimension formula, \Eq{em_dim},  
and inserting the values of the couplings
$g_{\alpha}$ from \Eq{couplingS},  we arrive at
\be{c_charge1}
c = 3 - 6\left( \frac{e_{\rm b}^2}{1 - e_{\rm b}}
      + \frac{e_{\rm g}^2}{1 - e_{\rm g}} \right),
\ee
where we recall that $n_{\rm b} = 2 \cos(\pi e_{\rm b})$ and similarly
for $n_{\rm g}$. In Table~\ref{Tab:c} the numerically calculated values of 
the conformal charge are compared to the above formula, and 
excellent agreement is found.

\section{Geometrical scaling dimensions} 
\label{sec_exp}

\subsection{Two-string dimension}

In addition to the central charge, the Coulomb gas representation of
the loop model provided by the Liouville field theory, \Eq{action_tot}, 
allows for the evaluation of various geometrical scaling dimensions. 
As an example of such a quantity, consider the
probability $G_2(r)$ that two points separated by a distance $r$ lie
on the same, say, black loop. In the critical phase we expect this 
probability to decay as $G_2(r) \sim r^{-2x_2}$, which defines  the 
scaling dimension $x_2$. Since a black loop  is  represented as a
sequence of alternating ${\bf A}$ and ${\bf B}$-coloured 
edges it follows from the 
colouring constraint that the microscopic heights ${\bf z}$ just
outside this loop differ by integer multiples of ${\bf C}$ and
${\bf D}$ only. In other words, a black loop is a {\em contour} loop
for the component of the height along the direction perpendicular to
both ${\bf C}$ and ${\bf D}$, i.e., the $(1,0,-1)$
direction in height space. 
Similarly the grey loops are contour loops for the height
component along the $(1,0,1)$ direction.

It has been argued that the scaling dimension governing the
probability that two points belong to the same contour loop of a
random Gaussian surface equals $1/2$, independent of the
stiffness \cite{jk_prl95}. Thus, for $(n_{\rm b},n_{\rm g}) = (2,2)$ 
when ${\bf e}_0={\bf 0}$ and the effective field 
theory is Gaussian, we expect  $x_2 = 1/2$. For other 
values of the fugacities the Gaussian theory is modified by the 
background charge and the same argument cannot be made. 

\begin{figure} 
 \centering\epsfig{file=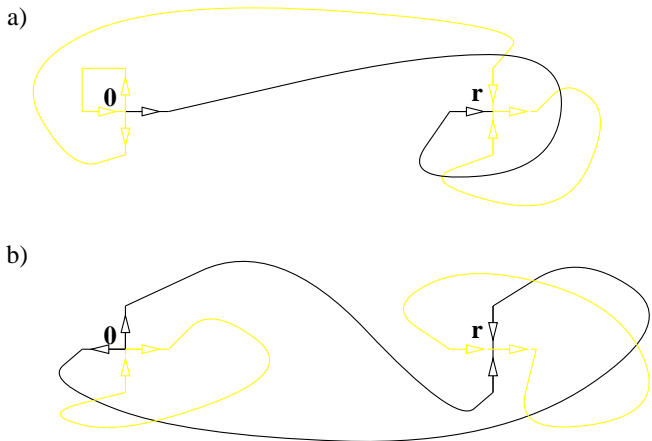,width=0.68\linewidth,angle=270}
 \begin{minipage}[t]{8.66cm}
 \caption{Defect configurations used to calculate geometrical exponents 
  $x_1$ (a) and $x_2$ (b) in the \F2 model. 
  In (a) there is a single oriented black 
  loop segment and a single oriented grey loop segment propagating from  
  ${\bf 0}$ to ${\bf r}$, whilst in (b) there are two oriented black 
  loop segments between ${\bf 0}$ and ${\bf r}$.}  
 \label{Fig:defects} 
 \end{minipage}
\end{figure} 

A more illuminating way of making contact with the interface
representation is to view $G_2(r)$ as  a two-string 
correlation function associated with defect configurations where two  
black strings emanating from the origin
annihilate one another at a distant point ${\bf r}$; see \Fig{Fig:defects}b.
This can be accomplished by rewriting $G_2(r)$ as $Z(r)/Z$, where $Z$
is the partition function defined by \Eq{part2}, and $Z(r)$ is
similarly defined but with the summation restricted to those
configurations ${{\cal G}'}_r$ where an oriented black loop passes through the
points ${\bf 0}$ and ${\bf r}$. Now consider reversing the direction
of one half of the loop, so that instead of having one oriented loop
passing through ${\bf 0}$ and ${\bf r}$ we have two oriented loop
segments directed from ${\bf 0}$ to ${\bf r}$\cite{Nienhuis}.
This corresponds to the
introduction of defect configurations at these two points, where we
have violated the edge-colouring constraint. 
At ${\bf 0}$
we find  a $({\bf C},{\bf D},{\bf A},{\bf A})$ 
configuration of colours which in the height language corresponds to a 
vortex of strength
\be{m2_strength}
 {\bf m}_2 = {\bf A} - {\bf B} = (-2,0,2) \ .
\ee
The strength of the vortex (its Burgers charge)  at  ${\bf 0}$ 
is calculated as the total height change around ${\bf 0}$.  
Similarly, at {\bf r} we have the corresponding antivortex
$({\bf B},{\bf B},{\bf C},{\bf D})$ of strength $-{\bf m}_2$ as
illustrated in Fig.~\ref{Fig:defects}b.

In order to calculate $x_2$ for general values of the loop fugacities
we have to take into account the effect of the complex phase factors 
associated with oriented loops. 
Namely, when one or more, say, black strings are associated with 
a vortex-antivortex configuration,   
spurious phase factors $\exp (\pm i \pi e_{\rm b})$ 
will arise whenever a black  loop segment winds  around one of the 
vortex cores \cite{Nienhuis}; for example, 
in \Fig{Fig:defects}b one of the two black strings  winds once arround 
point ${\bf r}$. The spurious winding phase can be removed  by
inserting the vertex operator
$\exp(i {\bf e}_{\rm b} \cdot {\bf h})$ at the positions of both vortex cores. 
Since a black loop has alternating ${\bf A}$ and ${\bf B}$ colours the  
electric charge ${\bf e}_{\rm b}$ must satisfy
\bea{eb-eqs} 
  {\bf e}_{\rm b} \cdot {\bf A} =  \pi e_{\rm b}, \ \ \ \  & 
 {\bf e}_{\rm b} \cdot {\bf B}  =  - \pi e_{\rm b}, \nonumber \\ 
 {\bf e}_{\rm b} \cdot {\bf C}  =   0, \ \ \ \ &
 {\bf e}_{\rm b} \cdot {\bf D}  =   0   \ . 
\eea
Similarly, if there are grey strings propagating between two vertices 
the spurious phase factors associated with winding configurations 
are corrected 
with vertex operators whose electric charge ${\bf e}_{\rm g}$ is 
determined by
\bea{eg-eqs} 
 {\bf e}_{\rm g} \cdot {\bf A}  =  0, \ \ \ \ &
 {\bf e}_{\rm g} \cdot {\bf B}  =  0, \nonumber \\ 
 {\bf e}_{\rm g} \cdot {\bf C}  =   \pi e_{\rm g},  \ \ \ \ & 
 {\bf e}_{\rm g} \cdot {\bf D}  =  -\pi e_{\rm g}   \ . 
\eea
Using \Eq{colour-vec} for the colour-vectors we find, 
\be{eg_eb} 
 {\bf e}_{\rm b} = -\frac{\pi}{2} (e_{\rm b},0,-e_{\rm b}), \ \ \ \
 {\bf e}_{\rm g} = -\frac{\pi}{2} (e_{\rm g},0,e_{\rm g}) \ . 
\ee
Going back to the two-string operator we conclude that it has 
total electromagnetic charge $({\bf e}_{\rm b},{\bf m}_2)$.

Finally, from the general expression for the scaling dimension of an 
electro-magnetic operator, \Eq{em_dim}, it follows that
\be{x2}
 2 x_2 = 2x({\bf e}_{\rm b},{\bf m}_2)
       = (1-e_{\rm b}) - \frac{e_{\rm b}^2}{1-e_{\rm b}}.
\ee
In Table~\ref{Tab:x2} exact values of $x_2$ calculated from this 
formula are compared to numerical results, and excellent agreement is found. 

Interestingly the expression for $x_2$  is {\em independent} of $e_{\rm g}$, 
i.e., it is not affected by the fugacity of grey loops. 
This observation conforms to our understanding of the scaling of 
compact polymers. 
The compact polymer problem is recovered in the limit $n_{\rm b}\to 0$ 
in which case there is a single black loop on the lattice. Since the loop
fills space its Hausdorff dimension is necessarily $D=2$. Scaling tells us 
that\cite{saleur_dupl87} 
\be{Haus_dim}
 D = 2 - x_2  
\ee
from which the result $x_2=0$ follows, {\em independent} of the fugacity 
of grey loops. The fact that our formula reproduces this simple result
in the $n_{\rm b}=0$ ($e_{\rm b}=1/2$) case provides a non-trivial check
on its validity. 

\subsection{One-string dimension}

The scaling dimension $x_1$ corresponding to
one black and one grey string propagating between two points on the 
lattice, can be computed in a way that is completely analogous to the 
case of two black strings discussed above.
(Note that the fully packing constraint ensures that if there
is a single black string between two points then these 
points are also connected by a grey string; see \Fig{Fig:defects}a.) 
Choosing one point on the 
even sublattice and the other on the odd, leads to the appearance of 
the defect configuration 
$({\bf A},{\bf C},{\bf C},{\bf D})$ on both sites of the square lattice. 
These in turn correspond to 
vortices in the height representation with topological charges
$\pm {\bf m}_1$, where   
\be{m1_strength}
 {\bf m}_1 = {\bf C} - {\bf B} = (-2,-2,0) \ .
\ee
Since strings of both flavours are now present the compensating 
electric charge
is ${\bf e}_{\rm b} + {\bf e}_{\rm g} = {\bf e}_0$. Hence
\bea{x1}
 2 x_1 &=& 2x({\bf e}_0,{\bf m}_1) \nonumber \\
       &=& \frac{1}{4} \left[ (1-e_{\rm b}) + (1-e_{\rm g}) \right] \\
       &+& \frac{(1-e_{\rm b})(1-e_{\rm g})}{(1-e_{\rm b})+(1-e_{\rm g})}
        -  \left[ \frac{e_{\rm b}^2}{1-e_{\rm b}} +
           \frac{e_{\rm g}^2}{1-e_{\rm g}} \right]. \nonumber
\eea
There are of course several different ways of choosing the defect
configurations (in this case, eight), but it should hardly come as a
surprise that they all lead to the same expression for the scaling
dimension. 

Unlike $x_2$, $x_1$ depends on both loop fugacities. Going back to our
original motivation,  the compact polymer problem 
($n_{\rm b}=0\Rightarrow e_{\rm b}=1/2$), 
$x_1$ determines the
value of the conformational exponent $\gamma= 1 - x_1$,  which 
describes the scaling of the number of compact polymers with the number of 
monomers. We see that depending on $e_{\rm g}$ there will be a 
continuum of $\gamma$'s. How do we interpret this? 

First note that the  problem of counting the number of 
conformations of a single compact polymer is the case $n_{\rm g}=1$
($e_{\rm g}=1/3$) which simply assigns equal weights to all conformations. 
Using \Eq{x1} this choice leads to $x_1=-5/112$ and 
to the result $\gamma=117/112$  advertised in the abstract.  
Changing $n_{\rm g}$ ($e_{\rm g}$) away from $n_{\rm g}=1$, on the other hand,
has the effect of favouring certain compact polymer conformations 
over others depending on the number of loops formed by the uncovered (grey)
bonds. In this sense the weight assigned to  grey loops  
can be thought of as an interaction between the 
monomers of the compact polymer, albeit a peculiar non-local one. 
Nonetheless, it is interesting that this interaction 
changes the scaling properties of the compact polymer leading to a 
continuously varying exponent $\gamma$ (more on this in the Discussion).

\subsection{Many-string dimensions}

The dimensions $x_1$ and $x_2$ given above are contained in a more
general set of string dimensions $x_{s_{\rb},s_{\rg}}$ 
governing the probability of 
having $s_{\rb}$ black loop segments and $s_{\rg}$ grey loop segments  
propagating between two points on the lattice \cite{duplantier_review}.
More precisely, we consider two microscopic regions centred
around points separated by a macroscopic distance, one region being the
source and the other the sink of the oriented loop segments. 
Since  the defect configurations obtained by
violations of the edge colouring constraint must necessarily give rise to
an {\em even} number of strings we will only consider the case when 
$s_{\rb}+s_{\rg}$ is even.  

Consider first the case $s_{\rb}=2k_{\rb}$ and $s_{\rg}=2k_{\rg}$. 
The appropriate magnetic charge is obtained by combining  
$k_{\rb}$ vortices with charge 
${\bf A}-{\bf B}=(-2,0,2)$, and $k_{\rg}$ vortices with charge 
${\bf C}-{\bf D}=(-2,0,-2)$.
The defect with charge ${\bf A}-{\bf B}$ acts as a source of two black 
segments, whilst ${\bf C}-{\bf D}$ is associated with two grey loop segments. 
We also need to introduce the electric charge $e_{\rb}+e_{\rg}$ to compensate 
for the extra winding phase associated with the black and grey loop segments. 
The total electromagnetic charge is therefore
\bea{em_even} 
  \lefteqn{ [{\bf e}_{2k_{\rb},2k_{\rg}}, {\bf m}_{2k_{\rb}, 2k_{\rg}}] =} \\
  & & \ \ \ \
  [{\bf e}_{\rb} (1-\delta_{k_{\rb},0})+{\bf e}_{\rg}(1-\delta_{k_{\rg},0}),
  -2(k_{\rb}+k_{\rg},0,k_{\rg}-k_{\rb})],  \nonumber
\eea
and from the dimension formula, \Eq{em_dim}, we find
\bea{x_even}
 \lefteqn{
 2 x_{2k_{\rb},2k_{\rg}} = } \nonumber \\ 
  & & \ \ \ \  (1-e_{\rm b}) k_{\rb}^2 + (1-e_{\rm g}) 
            k_{\rg}^2 +   \nonumber \\
  & & \ \ \ \  -   \frac{e_{\rm b}^2}{1-e_{\rm b}} (1-\delta_{k_{\rb},0})   
          -   \frac{e_{\rm g}^2}{1-e_{\rm g}} (1-\delta_{k_{\rg},0}).
\eea
This formula generalises \Eq{x2}. 

Similarly, for $s_{\rb}=2k_{\rb}-1$ and $s_{\rg}=2k_{\rg}-1$ 
the electromagnetic charge is
\bea{em_odd}
  \lefteqn{
  [{\bf e}_{2k_{\rb}-1,2k_{\rg}-1},{\bf m}_{2k_{\rb}-1,2k_{\rg}-1}] =} 
  \nonumber \\  
 & & \ \ \ \  [{\bf e}_0,-2(k_{\rb}+k_{\rg}-1, 1, k_{\rg}-k_{\rb})] ; 
\eea
the magnetic charge is obtained by combining $k_{\rb}-1$ defects of
charge ${\bf A}-{\bf B}$, $k_{\rg}-1$ defects of charge ${\bf C}-{\bf D}$, 
and a single defect of charge ${\bf C}-{\bf B}$ which produces the remaining  
single  black and grey strings originating from the same vertex. 
The scaling dimension is found to be
\bea{x_odd}
 \lefteqn{
 2 x_{2k_{\rb}-1, 2k_{\rg}-1} =} \nonumber \\ 
 & & \ \ \ \
 \frac{1}{4} \left[ (1-e_{\rm b}) (2k_{\rb}-1)^2 + (1-e_{\rm g})(2k_{\rg}-1)^2
                \right] +  \\ 
 & & \ \ \ \            
  \frac{(1-e_{\rm b})(1-e_{\rm g})}{(1-e_{\rm b})+(1-e_{\rm g})}
             -  \left[ \frac{e_{\rm b}^2}{1-e_{\rm b}} +
                \frac{e_{\rm g}^2}{1-e_{\rm g}} \right]. \nonumber
\eea
This generalises the expression given in Ref.~\cite{jk_prl} and
correctly reduces to \Eq{x1} for $k_{\rb},k_{\rg}=1$.

\subsection{Thermal dimension}

We now turn our attention to the thermal scaling dimension. The
FPL${}^2$ model can be thought of as the zero-temperature limit of a
more general model where we allow for thermal excitations that violate
the close packing constraint. In this sense the temperature variable
is thermodynamically conjugate to the constraint that every vertex be
visited by (say) a  black loop. An appropriate defect configuration for
computing $x_T$ within the FPL${}^2$ model is therefore
$({\bf C},{\bf D},{\bf C},{\bf D})$. This is a vortex of strength
\be{mT_strength}
 {\bf m}_T = 2 ({\bf C} + {\bf D}) = (0,-4,0),
\ee
and since no strings terminating in the bulk are generated there
is no compensating electric charge. The scaling dimension is
then 
\be{xT}
 2 x_T = 2 x(0,{\bf m}_T)
       = 4 \frac{(1-e_{\rm b})(1-e_{\rm g})}{(1-e_{\rm b})+(1-e_{\rm g})} \ .
\ee
The exact values of $x_T$ quoted in Table~\ref{Tab:xT} are calculated using 
this formula.  The numerical results are in   excellent 
agreement.

\subsection{Boundary-string dimensions} 

The simplest example of a string  operator that cannot be 
accessed within the formalism  presented above is that of
one black and no grey strings propagating between two vertices of the 
square lattice. Since this configuration has an odd number of strings 
connecting two sites of the lattice 
these two sites necessarily reside on the {\em boundary}. 

If we define the FPL${}^2$ model on the cylinder, as will be the case when we 
construct its transfer matrix in Sec.~\ref{sec_TM}, 
a single black string can be enforced to run along the length of the 
cylinder if its  circumference is chosen {\em odd}. Taking our cue from 
the formulae derived above for the bulk string operators we {\em guess} 
the  formula 
\be{x_twist}
  X = \frac{1}{8} + \frac{1-e_{\rm b}}{8} -
      \frac{1}{2} \frac{e_{\rm b}^2}{1-e_{\rm b}}
\ee
from the numerical results shown in Table~\ref{Tab:X}. $X$ is the 
scaling dimension of the boundary operator which corresponds to a
single black (or grey) string.
 
The Coulomb gas interpretation of the second and third term 
in \Eq{x_twist} is rather apparent  when one compares them to \Eq{x2}. 
The second term can be  rationalised  as coming from a magnetic charge 
$(-1,0,1)$ which is half the charge ${\bf m}_2$ in \Eq{m2_strength},  
associated with two black strings; 
this is saying that we have a partial dislocation  generated at the 
boundary. The third term is due to the compensating electric charge 
${\bf e}_{\rb}$ for a single black string, 
same as in the two-string case. 

The first, constant  term does not have an immediate interpretation. 
A possible scenario  is that it is due to the 
boundary condition imposed on the height by virtue of having a 
cylinder of odd circumference. Namely, a translation along the 
periodic coordinate by an amount equal to the circumference ($L$) 
exchanges an even site for an odd site (and {\em vice versa}) resulting in a 
transformation of the height: ${\bf h}(x^1,x^2)= {\rm P} {\bf h}(x^1+L, x^2)$. 
Since ${\rm P}^2=1$ this boundary condition can be thought of as an 
insertion of a twist operator into the partition function. The 
twist operator has  dimension $1/8$  regardless of the 
stiffness of the interface \cite{ginsparg}. 

The above considerations permit us to calculate the scaling dimension
for the general case of an odd number of strings. For definiteness we
consider the case of $s_{\rb} = 2 k_{\rb}-1$ and
$s_{\rg} = 2 k_{\rg}$. The magnetic charge pertaining to
this situation is found by combining $2k_{\rb}-1$ defects of charge
$\frac12 ({\bf A}-{\bf B})$ with $k_{\rm g}$ defects of charge
${\bf C} - {\bf D}$, totaling
\bea{em_odd_even}
  \lefteqn{
  [{\bf e}_{2k_{\rb}-1,2k_{\rg}},{\bf m}_{2k_{\rb}-1,2k_{\rg}}] =} \\  
 & & \ \ \ \  [{\bf e}_{\rb} + {\bf e}_{\rg}(1-\delta_{k_{\rm g},0}),
               (1-2k_{\rb}-2k_{\rg}, 0, 2k_{\rb}-2k_{\rg}-1)]. \nonumber
\eea
Taking into account the contribution from the twist operator, i.e., adding
$1/8$ to the result obtained from \Eq{em_dim},   the
scaling dimension is then
\bea{x_odd_even}
 \lefteqn{
 2 x_{2k_{\rb}-1, 2k_{\rg}} =} \nonumber \\ 
 & & \ \ \ \            
  \frac18 + 
  \frac{1}{4} \left[ (1-e_{\rm b}) (2k_{\rb}-1)^2 + (1-e_{\rm g})(2k_{\rg})^2
                \right] + \\ 
 & & \ \ \ \
  - \left[ \frac{e_{\rm b}^2}{1-e_{\rm b}} +
  \frac{e_{\rm g}^2}{1-e_{\rm g}}(1-\delta_{k_{\rg},0}) \right]. \nonumber
\eea

\subsection{Complete spectrum of string dimensions}

Finally, the results of Eqs.~(\ref{x_even}), (\ref{x_odd}) and
(\ref{x_odd_even}) can be combined into a single equation 
for the scaling dimension of a string operator that corresponds to 
$s_{\rb}$ black loop segments and $s_{\rg}$ grey loop segments:
\bea{x_general}
 \lefteqn{
 2 x_{s_{\rb}, s_{\rg}} =} \nonumber \\ 
 & & \ \ \ \            
  \frac18 \delta^{(2)}_{s_{\rb}+s_{\rg},1} +
  \frac{1}{4} \left[ (1-e_{\rm b}) s_{\rb}^2 + (1-e_{\rm g}) s_{\rg}^2
              \right] -  \\
 & & \ \ \ \ 
  \left[ \frac{e_{\rm b}^2}{1-e_{\rm b}}(1-\delta_{s_{\rb},0}) +
  \frac{e_{\rm g}^2}{1-e_{\rm g}}(1-\delta_{s_{\rg},0})
              \right] + \nonumber \\
 & & \ \ \ \ 
  \delta^{(2)}_{s_{\rb},1} \delta^{(2)}_{s_{\rg},1}
  \frac{(1-e_{\rm b})(1-e_{\rm g})}{(1-e_{\rm b})+(1-e_{\rm g})} \nonumber ; 
\eea
here  we have defined
$\delta^{(2)}_{i,j} \equiv \delta_{i=j \rm{(mod \ 2)}}$.

\section{Termination of critical behaviour}
\label{sec_term}

In the preceding sections we have developed an effective description of
the critical phase  of the FPL${}^2$ model in the form of a field theory. This
theory  has to break down at large values of the loop fugacity since 
in this case a  typical state of the model will consist of  small 
loops only, i.e., a power-law distribution of loop sizes will be 
absent. That this indeed happens can be seen from the Liouville field    
theory itself as it carries the seeds of its own demise.  

The mapping of the loop model to an oriented loop model for
$n_{\rm b}, n_{\rm g}\le 2$ works equally well for $n_{\rm b}>2$ or
$n_{\rm g}>2$. 
{}From \Eq{fug} 
it follows that in the latter case at least one of the parameters,
$e_{\rm b}$ or $e_{\rm g}$, will be pure imaginary. This affects  
the Liouville potential which for $n_{\rm b}>2$ or  $n_{\rm g}>2$ becomes a
{\em relevant} perturbation to the (modified) Gaussian action 
$S_{\rm E} + S_{\rm B}$. 

To understand how this comes about  we consider
the simple case provided by the $n_{\rm b}=n_{\rm g}$ FPL${}^2$ model, 
discussed in Sec.~\ref{loop_ansatz}. 
Namely, as we increase the value of the loop fugacity we  
expect small loops to be favoured and  the stiffness $K$ of the interface 
to grow. In the critical phase this is offset by the decrease 
in the background 
charge in a way that leaves the Liouville potential marginal. 
Now when the loop fugacity exceeds 2 the background charge
${\bf e}_0 = -\pi(e_{\rm b},0,0)$ becomes pure imaginary and
the dimension of the Liouville potential
\be{L_dim}
  x_{\rm L} = \frac{\pi}{2} \ \frac{1-e_{\rm b}}{K}
\ee
can no longer stay marginal;  here $x_{\rm L}\equiv 
x({\bf e}^{(i)}, 0)$, where the charges ${\bf e}^{(i)}$ are given in 
\Eq{marg_ch}, and their   
scaling dimensions are calculated from 
\Eq{em_dim2}. In fact, 
assuming that the stiffness $K$ continues to increase with 
the loop fugacity for $n_{\rb}=n_{\rg}>2$, 
$x_L$ turns  complex with a real part that 
is {\em smaller} than two, 
rendering the Liouville potential {\em relevant}.

If we make the usual assumption of no intervening fixed points,  the 
relevant Liouville 
potential will generate a finite correlation length and the loop model 
will no longer be critical. The correlation length has the physical 
interpretation of the average size of a loop in the system. 
This scenario has been confirmed for 
the fully packed loop model on the honeycomb lattice, from the 
Bethe ansatz solution of this model~\cite{kast}. 

A different view of the non-critical region of the \F2 model 
is provided by the locking potential
$V({\bf h})$. Namely, the discrete nature of the microscopic heights 
can be taken into account  in the field theory 
by a negative potential in height space that is peaked around 
the flat, ideal states. As such, this potential is uniform on the 
ideal state graph and can therefore be expanded in a Fourier series. 
Examination of the most relevant vertex operators in this series \cite{jk_prb}
reveals that they are the same as the ones for the loop-weight (Liouville) 
potential, 
$w({\bf h})$. Therefore, just like $w({\bf h})$, the locking potential 
in the non-critical region of the phase diagram is a {\em relevant} 
perturbation.  Thus, it will lock the interface in one of 
the ideal, flat states. In this flat phase the height fluctuations
are bounded (as opposed to being logarithmically divergent) which is
just another way
of saying that large contour loops are exponentially suppressed.
On the other hand,  in the critical region of the 
\F2 model the locking potential is marginal as it would be for an interface 
model {\em at} the roughening transition \cite{chui}. This might indicate that 
the whole critical region of the \F2 model 
can be understood as a manifold of essential 
singularities in some more general model, as was the case for the 
honeycomb FPL model \cite{jk_JPA,baxter3c}.

As advertised in Secs.~ \ref{sec_CC} and~\ref{sec_exp}  our
results for the central charge and a number of the geometrical scaling
dimensions have been very accurately confirmed by transfer matrix
calculations. Before turning to a discussion of our numerical results
we describe the particular representation of the transfer matrix used
to obtain them.

\section{Construction of the transfer matrix}
\label{sec_TM}

To construct the transfer matrix for the FPL${}^2$ model on a cylinder
of circumference $L$ we write the partition function as
\begin{equation}
  Z^{(M)} = \sum_{{\cal G}_M} n_{\rm b}^{N_{\rm b}} n_{\rm g}^{N_{\rm g}},
  \label{partition}
\end{equation}
where the length of the cylinder $M$ has been explicitly indicated.
Periodic boundary conditions are  imposed in the
horizontal direction, whereas the bottom and the top row of the cylinder
have open boundary conditions and hence terminate in $L$ dangling edges.
We recall that the restriction of the summation to the set of fully
packed graph configurations ${\cal G}_M$ implies that locally
the vertices are constrained to have one of the six appearances
shown in Fig.~\ref{Fig:vertices}. In the first four possible vertices the
loop segments do not cross, whilst in the last two vertices the two
flavours intersect. The global constraint that all loops be closed in the
limit of an infinite system means that loop segments cannot terminate in
the bulk but only at the dangling edges in the top and bottom rows.

\begin{figure}
 \centering\epsfig{file=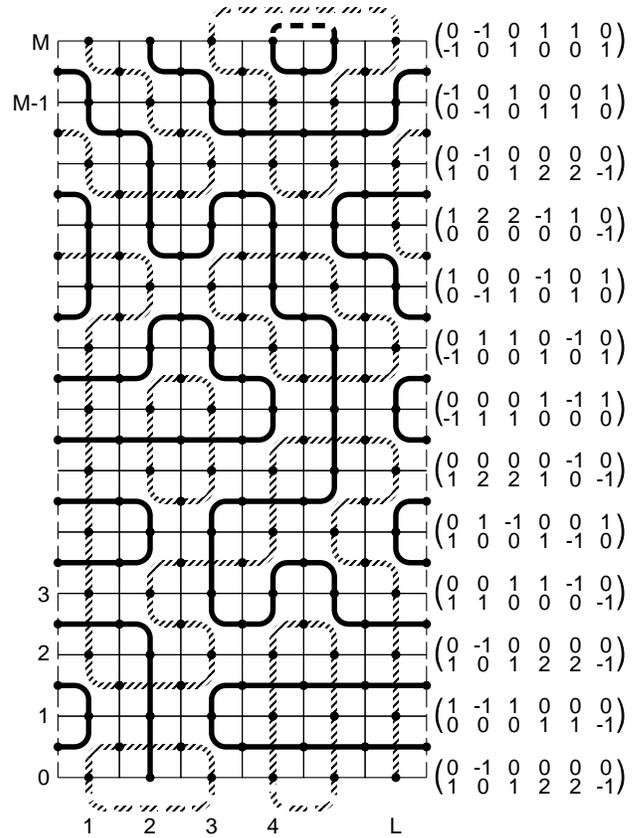,width=\linewidth,angle=0}
 \begin{minipage}[t]{8.66cm}
  \caption{A typical loop configuration for $L=6$ and $M=12$. The dashed
    lines along the left and the right boundaries illustrate the periodic
    boundary conditions. Horizontally the vertices are numbered from 1 to
    $L$, whilst vertically the rows are labeled by the number of the
    vertex immediately below them. This particular configuration is
    constrained to having precisely one string of each flavour spanning the
    length of the cylinder, and hence it contributes to the geometrical 
    exponent $x_1$. To the right we show the index representation of the
    connectivity state pertaining to each row (see Appendix
    \ref{app:Connect} for details).
    Any valid configuration can be interpreted as a ``jigsaw puzzle'' 
    assembled from
    the six ``pieces'' shown in Fig.~\ref{Fig:vertices}. Note that when
    laying down the first row of this puzzle it must be stipulated how the
    dangling edges of row 0, which are not part of a string, are pairwise
    interconnected below that row. These implicit connections as well as
    their counterparts in row $M$ have been depicted by dashed loop segments.}
  \label{Fig:lattice}
  \end{minipage}
\end{figure}

A typical loop configuration for a cylinder with $L=6$ and $M=12$ is shown
in Fig.~\ref{Fig:lattice}. The horizontal numbering pertains to the
vertices, whilst in the vertical direction it is more convenient to label
each row by the number of the vertex immediately below it. Accordingly the
labels 0 and $M$ refer to the bottom and the top row of dangling edges
respectively. We shall soon see that the inclusion in ${\cal G}_M$ of one
or more strings running between the dangling edges of row 0 and $M$ helps
us access the geometrical exponents of the model. In particular, the
configuration of Fig.~\ref{Fig:lattice} having one such string of each
flavour furnishes a contribution to the scaling 
dimension $x_1$ which determines the conformational  exponent $\gamma=1-x_1$.

\subsection{Connectivity basis}

The construction of a transfer matrix (TM) for \Eq{partition} appears to
be obstructed by the non-locality of  $N_i$ ($i = {\rm b}, {\rm g}$).
The key to solving 
this problem is to write the TMs in a basis of {\em connectivity states}
comprising information about how the dangling ends of row $M$ are pairwise
interconnected in the preceding rows and, if strings are present,
information about the positions of such strings. In addition the
connectivity states must keep track of the particular flavour of any loop
or string segment terminating in row $M$. Our construction generalises the
work of Bl\"{o}te and Nightingale for the $Q$-state Potts model
\cite{Blote82,Jesper3} and that of Bl\"{o}te and Nienhuis for the O($n$)
model \cite{Blote89} to take the extra flavour information into account,
and our notation is consistent with that of these authors.

It is essential to be able to represent a given connectivity state both in
an {\em index representation} giving direct access to the flavour and
connectedness information just mentioned, and in a {\em number representation} 
assigning an integer in the range $1,2,\ldots,C_L^{(s_{\rm b},s_{\rm g})}$
to the state under consideration. The latter representation enables us to
enumerate the entries of the TM, whilst the former allows us to determine
the number of loop closures when going from one connectivity state to
another and hence the value of a particular entry in the TM.
Here $C_L^{(s_{\rm b},s_{\rm g})}$ is the number of distinct
connectivity states for a cylinder of width $L$ accommodating $s_i$ strings
of flavour $i = {\rm b},{\rm g}$. The construction of these two
representations, the mapping between them, and the evaluation of the
$C_L^{(s_{\rm b},s_{\rm g})}$ for
$(s_{\rm b},s_{\rm g}) = (0,0)$, (1,0), (1,1) and (2,0) is
deferred to Appendix \ref{app:Connect}.

Designating the connectivity states by Greek letters we can write the
partition function as a sum of restricted partition functions
\begin{equation}
 Z^{(M)} = \sum_{\beta} Z_{\beta}^{(M)}
         = \sum_{\beta} \sum_{{\cal G}_M}
           \delta \big( \beta,\phi({\cal G}_M) \big)
           n_{\rm b}^{N_{\rm b}} n_{\rm g}^{N_{\rm g}},
\end{equation}
where $\phi({\cal G}_M)$ is the connectivity of the $L$ dangling edges of
row $M$, and $\delta (i,j)$ is the Kronecker delta.
Now consider appending another row to the cylinder, giving us a total graph
configuration  ${\cal G}_{M+1} = {\cal G}_M \cup {\cal G'}$.
Evidently the connectivity of the dangling edges of row $M+1$ is determined
solely by that of the preceding row and by the appended subgraph ${\cal G'}$
\begin{equation}
  \phi({\cal G}_{M+1}) = \psi \big( \phi({\cal G}_M),{\cal G'} \big).
\end{equation}
Letting $N'_i$ denote the number of loop closures induced by
${\cal G'}$ we arrive at the relation
\begin{eqnarray}
  Z_{\alpha}^{(M+1)} &=& \sum_{{\cal G}_{M+1}}
      \delta \big( \alpha,\phi({\cal G}_{M+1}) \big)
      n_{\rm b}^{N_{\rm b} + N'_{\rm b}}
      n_{\rm g}^{N_{\rm g} + N'_{\rm g}} \nonumber \\
  &=& \sum_{\beta} \sum_{{\cal G}_M} \delta \big( \beta,\phi({\cal G}_M) \big)
      n_{\rm b}^{N_{\rm b}} n_{\rm g}^{N_{\rm g}} \nonumber \\
  & & \sum_{{\cal G'} | {\cal G}_M}
      \delta \big( \alpha,\psi(\phi({\cal G}_M),{\cal G'}) \big)
      n_{\rm b}^{N'_{\rm b}} n_{\rm g}^{N'_{\rm g}} \nonumber \\
  &=& \sum_{\beta} T_{\alpha \beta} Z_{\beta}^{(M)},
\end{eqnarray}
where the transfer matrix is defined by
\begin{equation}
  T_{\alpha \beta} = \sum_{{\cal G'} | {\cal G}_M}
    \delta \big( \alpha,\psi(\phi({\cal G}_M),{\cal G'}) \big)
    n_{\rm b}^{N'_{\rm b}} n_{\rm g}^{N'_{\rm g}}.
  \label{transfer}
\end{equation}
The notation ${\cal G'} | {\cal G}_M$ means that the summation is
constrained to those subgraphs ${\cal G'}$ that fit the dangling
edges of ${\cal G}_M$.

\subsection{Single-vertex decomposition}

A quintessential step in the practical implementation of the TM is its
decomposition into matrices each corresponding to the addition of a single
vertex, 
\begin{equation}
  {\bf T} = {\bf T}_L \cdot {\bf T}_{L-1} \cdot \ldots \cdot {\bf T}_1.
\end{equation}
Here the single-vertex matrix ${\bf T}_i$, which adds the vertex at horizontal
position $i$ of the new row, has the advantage of being
{\em sparse}, and we shall soon see that it has at most three non-zero
entries per column. This property leads to a dramatical reduction of the
time and storage requirements for the calculations.

\begin{figure}
 \centering\epsfig{file=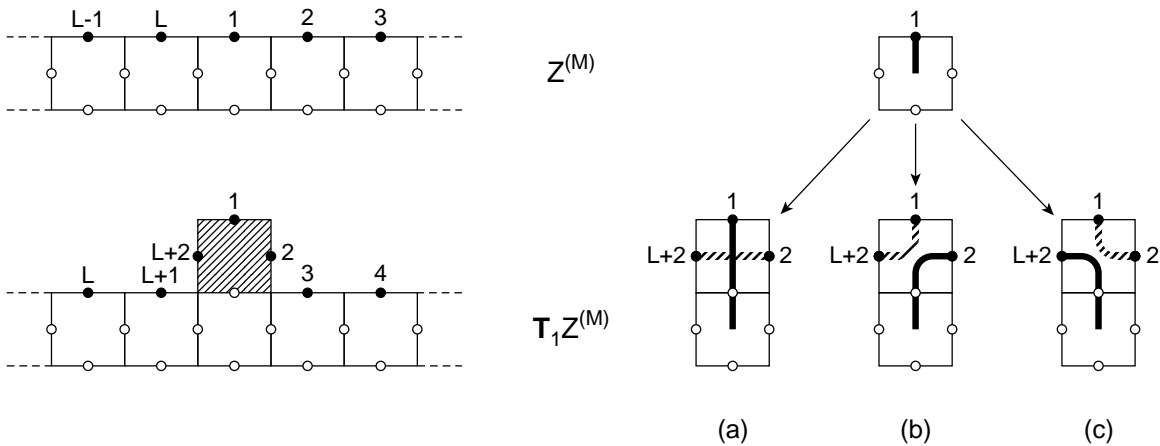,width=0.90\linewidth,angle=0}
  \caption{Adding the first vertex of the $(M+1)$'th row increases the
    number of dangling edges from $L$ to $L+2$. The labeling of the
    ``active'' edges (filled circles) before and after addition of the new
    vertex (shaded) is as shown in the 
    left part of the figure. The part of the lattice relevant for
    determining which of the vertices of Fig.~\ref{Fig:vertices} fit
    onto a given connectivity of row $M$, has been depicted in the right
    part of the figure. This information constitutes the {\em vertex rules},
    and is explained in the text.}
  \label{Fig:T1}
\end{figure}

As was the case in the O($n$) model \cite{Blote89}, a minor complication
arises due to the fact that the addition of the first vertex of a new row
increases the number of dangling edges from $L$ to $L+2$. This is
illustrated in the left part of Fig.~\ref{Fig:T1}. Upon addition of further
vertices the number of dangling edges is kept fixed at $L+2$, until the
$L$'th vertex completes the row, and we are back at $L$ dangling
edges. Thus the dimensions of the single-vertex matrices are
$C_{L+2} \times C_L$ for ${\bf T}_1$, $C_{L+2} \times C_{L+2}$ for
${\bf T}_2,\ldots,{\bf T}_{L-1}$, and $C_L \times C_{L+2}$ for ${\bf T}_L$.

\end{multicols}

\begin{multicols}{2}

In Fig.~\ref{Fig:T1} we illustrate the action of ${\bf T}_1$ on $Z^{(M)}$
in detail. To ensure that row $M+1$, when completed, will have the same
labels on its dangling edges as was the case in the preceding row, the 
solid dots illustrating the ``active'' dangling edges must be relabeled as
shown in the lower left part of the figure. Shown to the right are the three
possible choices of vertices fitting onto a black loop segment
terminating at the dangling end 1 of $Z^{(M)}$. There are thus three
non-zero entries in each column of ${\bf T}_1$. Since no loop closures of
either flavour can be induced ($N_{\rm b}'=N_{\rm g}'=0$ in \Eq{transfer}) all
these entries are unity. Similar considerations hold true when the loop
segment to be fitted is grey, and the vertex rules can be read off
from the figure by interchanging the two flavours.

\end{multicols}

\begin{figure}
  \centering\epsfig{file=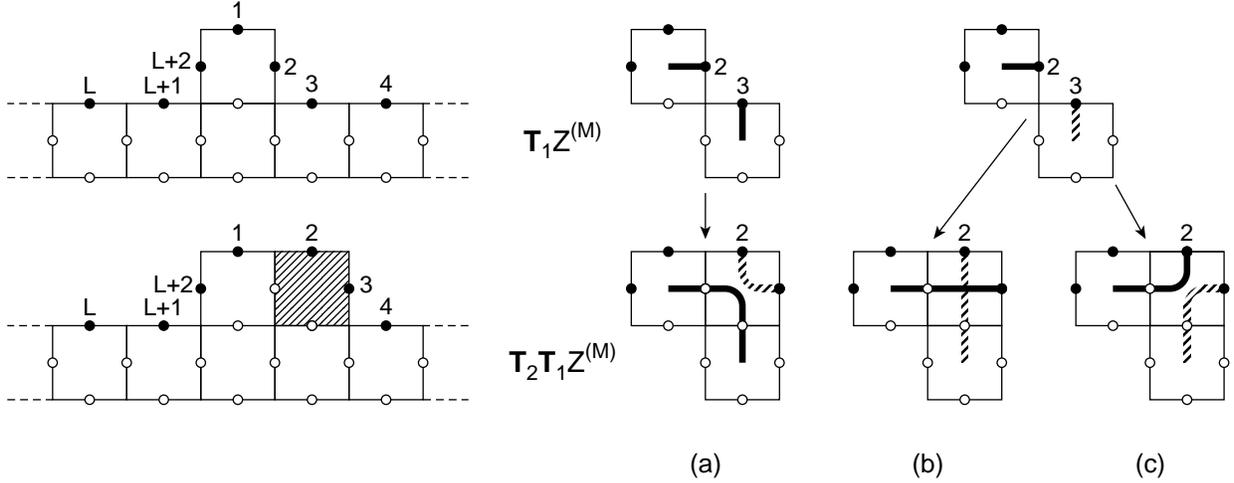,width=0.95\linewidth,angle=0}
  \caption{Addition of subsequent vertices keeps the number of dangling
    edges fixed at $L+2$. In the left part of the figure the system is
    shown before and after the addition of the second vertex (shaded). Vertex
    rules are displayed to the right. Situation (a) allows for the
    possibility of a black loop closure.}
  \label{Fig:T2}
\end{figure}

\begin{multicols}{2}

When acting with any one of the subsequent single-vertex TMs
${\bf T}_2,\ldots,{\bf T}_{L-1}$ the situation is as depicted in
Fig.~\ref{Fig:T2} for the case of ${\bf T}_2$. As the number of dangling
edges is kept fixed no relabeling is needed, apart from the translation of
labels 2 and 3 up on top of the newly added vertex. The vertex rules for
the case where edge 2 of ${\bf T}_1 Z^{(M)}$ is black are
shown in the right part of the figure; similar rules for the case where it
is grey can be obtained by permuting  the two flavours.

In situation (a) only one vertex fits onto the two dangling edges. The
column of ${\bf T}_2$ determined by the number representation of the
connectivity pertaining to the $L+2$ dangling ends that are active in the
upper part of the figure thus has only one non-zero entry. Its value is
either $n_{\rm b}$ or 1 depending on whether a black loop closure is induced
($i^{\rm b}_2 = i^{\rm b}_3$) or not ($i^{\rm b}_2 \neq i^{\rm b}_3$).
In the index representation of the new connectivity state
$i^{\rm g}_2 = i^{\rm g}_3$ is set equal to a positive
integer not assumed by any other $i^{\rm g}_k$. The new values of the black
indices depend on whether a loop closure is induced or not. In the former
case we simply set $i^{\rm b}_2 = i^{\rm b}_3 = 0$. In the latter, the
two left-over black partners must be mutually connected before assigning
$i^{\rm b}_2 = i^{\rm b}_3 = 0$.

Situations (b) and (c) correspond to two entries of each column of
${\bf T}_2$ taking the value unity, the others being zero. Since loop
closures are out of the question the handling of these cases is simple. In
(b) the two flavours cross, and the indices of sites 2 and 3 are interchanged. 
Case (c) is even simpler: it corresponds to a diagonal entry in ${\bf T}_2$.

\end{multicols}

\begin{figure}
  \centering\epsfig{file=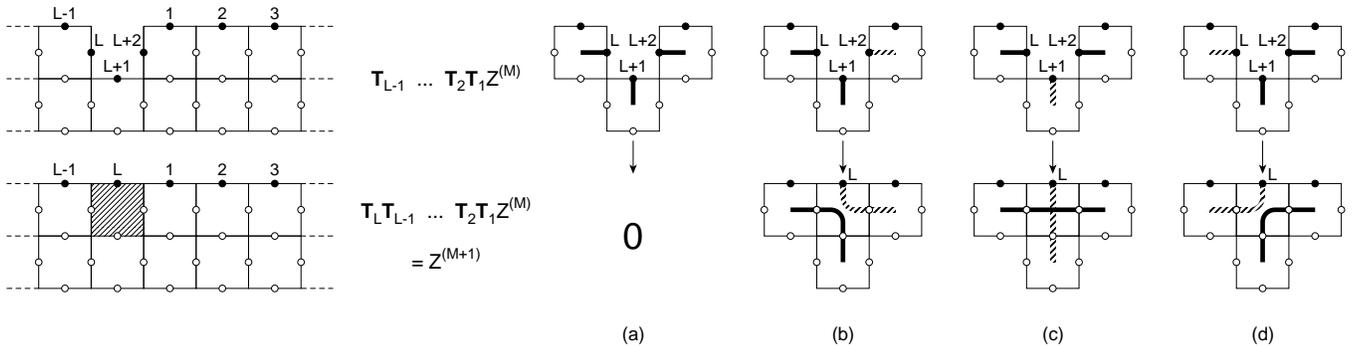,width=\linewidth,angle=0}
  \caption{Completing the $(M+1)$'th row by adding the $L$'th vertex
    (shaded) brings the number of dangling edges back to $L$. The labeling
    is now consistent with that of the preceding rows. Vertex rules, shown
    to the right, now include a disallowed configuration. Namely, in
    situation (a) none of the vertices of Fig.~\ref{Fig:vertices} fit
    in, and the corresponding entry of the transfer matrix must be forced
    to zero. Situations (b), (c) and (d) offer various possibilities
    for a black loop closure.}
  \label{Fig:TL}
\end{figure}

\begin{multicols}{2}

When strings are present a few modifications of the above rules are 
necessary.   In situation (a), if one of $i^{\rm b}_2$ and $i^{\rm b}_3$
equals -1 and the other is positive, the left-over partner to the
non-string black segment must be made the new string. And if both
$i^{\rm b}_2$ and $i^{\rm b}_3$ equal
-1 the corresponding entry of ${\bf T}_2$ must be forced to zero, since two
strings cannot be allowed to annihilate.

Finally, consider closing the $(M+1)$'th row through the action of
${\bf T}_L$, as depicted in Fig.~\ref{Fig:TL}. The labels $L+1$ and $L+2$
now disappear, and as far as the labeling goes the system is back in its
original state. Each column of ${\bf T}_L$ has at most one non-zero entry per
column, as witnessed by the vertex rules displayed in the right part of the
figure. Once again, only half of the vertex rules are shown, and the other half
is found by interchanging the two flavours. 

In situation (a) no vertex of Fig.~\ref{Fig:vertices} can fit onto the
three dangling edges at positions $L$, $L+1$ and $L+2$. The corresponding
entry of ${\bf T}_L$ must therefore be forced to zero. Situations (b), (c)
and (d) leave us to determine whether, for a given connectivity of the
$L+2$ dangling edges, a black loop closure occurs or not. The handling
in terms of the index representation is exactly as described above.

\section{Numerical results}
\label{sec_num}

\subsection{Central charge}

The reduced free energy per vertex in the limit $M \rightarrow \infty$ of an
infinitely long cylinder is given by
\begin{equation}
  f_0^{(0,0)}(L) = \lim_{M \rightarrow \infty} \frac{1}{LM}
                   \ln {\rm Tr} \ Z^{(M)}
                 = - \frac{1}{L} \ln \lambda_0^{(0,0)},
\end{equation}
where $\lambda_0^{(s_{\rm b},s_{\rm g})}$ is the largest eigenvalue of
${\bf T}^{(s_{\rm b},s_{\rm g})}$. The partition function for a
cylinder of length $M$ is found by iterating the no-string TM
\begin{equation}
  Z^{(M)} = \left( {\bf T}^{(0,0)} \right)^M Z^{(0)}.
\end{equation}

It is well-known that conformal invariance relates the amplitude of the
$1/L^2$ corrections to $f_0^{(0,0)}(\infty)$ to the central charge $c$
\cite{Cardy86}. A further (non-universal) $1/L^4$ correction due to the
operator $T \overline{T}$, where $T$ denotes the stress tensor, must also be
present in any conformally invariant system \cite{Cardy-CFT}. It is
therefore found in a number of cases \cite{Blote82,Queiroz95,Jesper4} that
fits of the form
\begin{equation}
  f_0^{(0,0)}(L) = f_0^{(0,0)}(\infty) - \frac{\pi c}{6L^2} + \frac{A}{L^4}
  \label{FSS-c}
\end{equation}
yield very rapidly converging estimates for $c$.
An efficient application of \Eq{FSS-c} is to determine $c$ from
parabolic least-squares fits of the finite-size data against $1/L^2$
\cite{Queiroz95,Jesper4}.

In Table \ref{Tab:c} the results of such fits including the data points
for $L_0 \le L \le L_{\rm max}$ are  shown as a function of $L_0$.
Numerically we were able to access $L_{\rm max} = 14$, in which case the
largest single-vertex TMs have dimension $\sim 7 \cdot 10^6$ (see Table
\ref{Tab:connect}).
The extrapolation of the estimants $c(L_0,L_{\rm max})$ to the limit of
infinite $L_0$ is assumed
to take the form of a power law
\begin{equation}
  c(L_0,L_{\rm max}) = c + k L_0^{-p},
\end{equation}
at least within an asymptotic regime of large enough $L_0$. As is evident
from Table \ref{Tab:c} the last three estimants usually exhibit
monotonicity, thus allowing us to fix the constants $c$, $k$ and $p$.
When this was not the case, or whenever the power $p$ thus obtained
was to small to produce a reliable extrapolation the Ising-like value
$p=2$ was used by default to extrapolate the last two estimants.
An error bar for this type of fit can be estimated from the variation
among the individual estimants. The extrapolants are invariably in
excellent agreement with our analytical results, the relative deviation
being typically of the order $10^{-3}$.

\end{multicols}

\begin{table}
 \begin{tabular}{rrrrrrrr}
    $n_{\rm b}$ &
    $n_{\rm g}$ & $c(4,14)$  & $c(6,14)$     & $c(8,14)$
                & $c(10,14)$ & Extrapolation & Exact           \\ \hline
    0.0 &   0.0 &    -2.8943 &       -2.8861 &       -2.9220
                &    -2.9514 &       -3.0037 &       -3.0000   \\ \hline
    0.5 &   0.0 &    -1.8528 &       -1.7641 &       -1.7716
                &    -1.7873 &       -1.8152 &       -1.8197   \\
    0.5 &   0.5 &    -0.7295 &       -0.6249 &       -0.6159
                &    -0.6220 &       -0.6328 &       -0.6395   \\ \hline
    1.0 &   0.0 &    -1.0012 &       -0.9542 &       -0.9636
                &    -0.9761 &       -0.9983 &       -1.0000   \\
    1.0 &   0.5 &     0.1341 &        0.1877 &        0.1924
                &     0.1895 &        0.1843 &        0.1803   \\
    1.0 &   1.0 &     0.9918 &        0.9969 &        0.9986
                &     0.9999 &        1.0004 &        1.0000   \\ \hline
    1.5 &   0.0 &    -0.3765 &       -0.3669 &       -0.3817
                &    -0.3923 &       -0.4111 &       -0.4124   \\
    1.5 &   0.5 &     0.7652 &        0.7746 &        0.7729
                &     0.7715 &        0.7690 &        0.7678   \\
    1.5 &   1.0 &     1.6215 &        1.5818 &        1.5778
                &     1.5806 &        1.5856 &        1.5876   \\
    1.5 &   1.5 &     2.2541 &        2.1691 &        2.1581
                &     2.1627 &        2.1709 &        2.1751   \\ \hline
    2.0 &   0.0 &     0.0706 &        0.0549 &        0.0342
                &     0.0235 &       -0.0019 &        0.0000   \\
    2.0 &   0.5 &     1.2209 &        1.1937 &        1.1868
                &     1.1861 &        1.1849 &        1.1803   \\
    2.0 &   1.0 &     2.0792 &        2.0002 &        1.9899
                &     1.9937 &        2.0005 &        2.0000   \\
    2.0 &   1.5 &     2.7139 &        2.5919 &        2.5737
                &     2.5781 &        2.5859 &        2.5876   \\
    2.0 &   2.0 &     3.1629 &        3.0121 &        2.9885
                &     2.9936 &        3.0027 &        3.0000   \\
 \end{tabular}
 \caption{Estimants $c(L_0,L_{\rm max})$ for the central charge are
   obtained from parabolic least-squares fits against $1/L^2$ using the
   numerical data for  $L_0 \le L \le L_{\rm max}$. The extrapolation
   in $L_0$ is described in the text.}
 \label{Tab:c}
\end{table}

\begin{multicols}{2}

The results for $c$ are shown for all integer and half-integer values of 
$n_i \in [0,2]$. Because of the symmetric appearance of the two flavours in
\Eq{partition} only $n_{\rm b} \ge n_{\rm g}$ need be considered. For either
$n_{\rm b} = 1$ or $n_{\rm g} = 1$
the FPL$^2$ model reduces to the simpler FPL model earlier considered by
Batchelor {\em et al.} \cite{Batchelor96}, and for $n_{\rm b} = n_{\rm g}$ we
recover another special case recently investigated by one of us \cite{jk_prl}.

\subsection{Thermal scaling dimension}

A further prediction of conformal invariance is that the finite-size scaling
of the first gap in the eigenvalue spectrum of ${\bf T}^{(0,0)}$ is related
to the thermal scaling dimension \cite{Cardy83}
\begin{equation}
  f_1^{(0,0)}(L) - f_0^{(0,0)}(L) = \frac{2 \pi x_T}{L^2} + \cdots,
  \label{FSS-x}
\end{equation}
where $f_1^{(0,0)}$ is found from the next-largest eigenvalue of
${\bf T}^{(0,0)}$ through $f_1^{(0,0)} = - \frac{1}{L} \ln \lambda_1^{(0,0)}$.
These computations were also carried through for even $L$ up to
$L_{\rm max} = 14$. In this case as well the convergence of the
estimants can be considerably sped up by including a $1/L^4$ term in
\Eq{FSS-x} and performing parabolic least-squares fits versus $1/L^2$.

The results for $x_T$ as displayed in Table \ref{Tab:xT} again agree with
those of the previously studied special cases \cite{Batchelor96,jk_prl}.
The data for $(n_{\rm b},n_{\rm g})=(0,0)$ merit a special comment.
Monitoring the
three leading eigenvalues $\lambda_0^{(0,0)}$, $\lambda_1^{(0,0)}$ and
$\lambda_2^{(0,0)}$ as a function of $n$ for $n_{\rm b} = n_{\rm g} \equiv n$
we found that  
$\lambda_1^{(0,0)}$ and $\lambda_2^{(0,0)}$ are exactly degenerate for all
$n$ down to $n \sim 0.20$. Hereafter $\lambda_1^{(0,0)}$ splits off from
$\lambda_2^{(0,0)}$ and eventually becomes degenerate with
$\lambda_0^{(0,0)}$ at $n=0$. Because of this level crossing it thus seems
very likely that near $(n_1,n_2) = (0,0)$ the thermal eigenvalue should be
related to the gap $f_2^{(0,0)}(L) - f_0^{(0,0)}(L)$. Comparison with the
exactly known result $x_T = 1/2$ \cite{jk_prl} confirms this
suspicion. A similar comment holds true near $(n_{\rm b},n_{\rm g}) = (2,2)$,
and again we find fair agreement with the exact result if we apply \Eq{FSS-x}
to $\lambda_2^{(0,0)}$, and not to $\lambda_1^{(0,0)}$ (which in this case 
becomes two-fold degenerate).

\end{multicols}

\begin{table}
 \begin{tabular}{lllllllll}
     $n_{\rm b}$ &
     $n_{\rm g}$ & $x_T(4,14)$ & $x_T(6,14)$ & $x_T(8,14)$ & $x_T(10,14)$
        & Extrapolation        & Ref.~[14]                 & Exact    \\ \hline
    0.0 &   0.0  &     0.5712  &     0.5280  &     0.5121  &  0.5060
        &              0.4987  &                           &  0.5000  \\ \hline
    0.5 &   0.0  &     0.5704  &     0.5535  &     0.5452  &  0.5417
        &              0.5366  &                           &  0.5372  \\
    0.5 &   0.5  &     0.5916  &     0.5882  &     0.5845  &  0.5825
        &              0.5789  &                           &  0.5804  \\ \hline
    1.0 &   0.0  &     0.5826  &     0.5798  &     0.5765  &  0.5748
        &              0.5708  &     0.573 (1)             &  0.5714  \\
    1.0 &   0.5  &     0.6204  &     0.6227  &     0.6218  &  0.6211
        &              0.6199  &     0.6200 (5)            &  0.6206  \\
    1.0 &   1.0  &     0.66368 &     0.66600 &     0.66642 &  0.66654
        &              0.66663 &     0.6666 (1)            &  0.66667 \\ \hline
    1.5 &   0.0  &     0.5965  &     0.6053  &     0.6060  &  0.6058
        &              0.6054  &                           &  0.6063  \\
    1.5 &   0.5  &     0.6493  &     0.6559  &     0.6574  &  0.6578
        &              0.6585  &                           &  0.6619  \\
    1.5 &   1.0  &     0.7782  &     0.7094  &     0.7108  &  0.7115
        &              0.7130  &     0.713 (1)             &  0.7146  \\
    1.5 &   1.5  &     0.8950  &     0.7657  &     0.7674  &  0.7684
        &              0.7702  &                           &  0.7699  \\ \hline
    2.0 &   0.0  &     0.6167  &     0.6295  &     0.6338  &  0.6349
        &              0.6356  &                           &  0.6667  \\
    2.0 &   0.5  &     0.7481  &     0.6878  &     0.6913  &  0.6927
        &              0.6945  &                           &  0.7345  \\
    2.0 &   1.0  &     0.8741  &     0.7566  &     0.7552  &  0.7565
        &              0.7588  &     0.76 (1)              &  0.8000  \\
    2.0 &   1.5  &     0.9436  &     0.8755  &     0.8284  &  0.8303
        &              0.8337  &                           &  0.8702  \\
    2.0 &   2.0  &     0.9996  &     0.9850  &     0.9400  &  0.9200
        &              0.8876  &                           &  1.0000  \\
 \end{tabular}
 \caption{The thermal scaling dimension $x_T$. The extrapolation of the
   estimants $x_T(L_0,L_{\rm max})$ is described in the text. For
   comparison we also show the numerical data for the
   case of either $n_{\rm b}$ or $n_{\rm g}$ being unity [14].
   Due to level crossing the values of $x_T$ for
   $(n_{\rm b},n_{\rm g}) = (0,0)$ and $(2,2)$ are found from the gap
   $f_2^{(0,0)}(L) - f_0^{(0,0)}(L)$ rather than from
   $f_1^{(0,0)}(L) - f_0^{(0,0)}(L)$.}
 \label{Tab:xT}
\end{table}

\begin{multicols}{2}

For $n_{\rm b} < 2$ the extrapolants are again in excellent ($\sim 10^{-3}$
or better) agreement with our analytical results. For $n_{\rm b} = 2$ the
slower convergence can be attributed to logarithmic corrections
\cite{Cardy-log}
arising from an enhanced number of marginal vertex operators. Indeed, of the
twelve vertex operators corresponding to the shortest vectors in 
${\cal R}^*_w$, \Eq{F_sum2}, 
seven stay marginal when either $n_{\rm b} < 2$ or $n_{\rm g} < 2$. 
In the general case, when both $n_{\rm b} < 2$ and $n_{\rm g} < 2$, there are
only four marginal vertex operators; this is the loop ansatz, \Eq{marg_dim}.

\subsection{Dimensions of string operators}

We now turn our attention to the determination of the scaling dimensions
associated with one or more strings spanning the length of the
cylinder. The presence of one black string corresponds to a height
mismatch in the ideal states, and the relevant scaling dimension $X$ is
therefore that of a {\em twist-like operator} \cite{ginsparg}. 
We have calculated the leading
eigenvalue of ${\bf T}^{(1,0)}$ for odd system sizes up to $L_{\rm max} = 13$
and determined the corresponding estimants $c_{\rm odd}(L_0,L_{\rm max})$
by the usual parabolic fits to $f_0^{(1,0)}(L)$, cfr.~\Eq{FSS-c}.
Estimants $X(L_0,L_{\rm max})$ are then defined by
\begin{equation}
  X(L_0,L_{\rm max}) =
  \frac{c - c_{\rm odd}(L_0,L_{\rm max})}{12},
\end{equation}
where the factor of 12 originates from a comparison of \Eq{FSS-c}
with \Eq{FSS-x}. For the central charge $c$ of an even-sized
system we use our analytical results, \Eq{c_charge1}.

These estimants and their extrapolations are found in Table
\ref{Tab:X}. Note that we can no longer limit the parameter values by
$n_{\rm b} \ge n_{\rm g}$, as the condition $(s_{\rm b},s_{\rm g}) = (1,0)$
treats the two flavours asymmetrically. In the case of the FPL model
($n_{\rm b} = 1$) it was found \cite{Batchelor96} that $X$ was
independent of $n_{\rm g}$. It is evident from
our numerical data that this $n_{\rm g}$-independence in fact pertains to all
$n_{\rm b} \in [0,2]$. Final results for $X$ as a function of $n_{\rm b}$ have
therefore been computed by averaging the available extrapolated scaling
dimensions over $n_{\rm g}$. For $n_{\rm b} = 1$ the agreement with
the result $X \approx 1/8$ 
found by Batchelor {\em et al.} \cite{Batchelor96} is excellent.
Furthermore we are able to conjecture the general formula,
\Eq{x_twist}, for $X$ as a function of the loop fugacities.

When $(s_{\rm b},s_{\rm g}) = (1,1)$ the parity of $L$ must again be
even, and we can
make parabolic fits for the gap $f_0^{(1,1)}(L) - f_0^{(0,0)}(L)$, as in
\Eq{FSS-x}, without taking resort to the less accurate method of
fitting for two central charges separately as above. The corresponding
universal amplitude is identified with the scaling dimension $x_1$. The
results, now for $L_{\rm max} = 12$, are shown in Table \ref{Tab:x1},
and our values for the scaling dimension are once again in agreement
with the analytical results, apart from $n_{\rm g} = 2$ 
where logarithmic corrections are the most likely source of systematic
errors \cite{Cardy-log}. 

\end{multicols}

\begin{table}
 \begin{tabular}{rrrrrrrrr}
    $n_{\rm b}$ &
    $n_{\rm g}$ & $X(3,13)$ & $X(5,13)$ & $X(7,13)$ & $X(9,13)$
        & Extrapolation     & Result                & Exact      \\ \hline
    0.0 &   0.0 &  -0.05586 &  -0.06109 &  -0.06203 &   -0.06232
        &          -0.06257 &  -0.06269 (31)        &   -0.06250 \\
    0.0 &   0.5 &  -0.06080 &  -0.06197 &  -0.06220 &   -0.06233
        &          -0.06253 &                       &   -0.06250 \\
    0.0 &   1.0 &  -0.06043 &  -0.06198 &  -0.06221 &   -0.06233
        &          -0.06250 &                       &   -0.06250 \\
    0.0 &   1.5 &  -0.05869 &  -0.06156 &  -0.06215 &   -0.06233
        &          -0.06259 &                       &   -0.06250 \\
    0.0 &   2.0 &  -0.05804 &  -0.06190 &  -0.06297 &   -0.06316
        &          -0.06324 &                       &   -0.06250 \\ \hline
    0.5 &   0.0 &   0.04674 &   0.04538 &   0.04558 &    0.04569
        &           0.04587 &   0.04583 (16)        &    0.04591 \\
    0.5 &   0.5 &   0.04572 &   0.04585 &   0.04589 &    0.04588
        &           0.04588 &                       &    0.04591 \\
    0.5 &   1.0 &   0.04643 &   0.04622 &   0.04614 &    0.04607
        &           0.04595 &                       &    0.04591 \\
    0.5 &   1.5 &   0.04781 &   0.04675 &   0.04638 &    0.04622
        &           0.04590 &                       &    0.04591 \\
    0.5 &   2.0 &   0.04828 &   0.04664 &   0.04593 &    0.04573
        &           0.04555 &                       &    0.04591 \\ \hline
    1.0 &   0.0 &   0.11895 &   0.12278 &   0.12398 &    0.12438
        &           0.12501 &   0.12497 (8)         &    0.12500 \\
    1.0 &   0.5 &   0.12346 &   0.12422 &   0.12458 &    0.12470
        &           0.12489 &                       &    0.12500 \\
    1.0 &   1.0 &   0.12465 &   0.12485 &   0.12495 &    0.12496
        &           0.12498 &                       &    0.12500 \\
    1.0 &   1.5 &   0.12584 &   0.12540 &   0.12529 &    0.12521
        &           0.12508 &                       &    0.12500 \\
    1.0 &   2.0 &   0.12652 &   0.12549 &   0.12513 &    0.12501
        &           0.12490 &                       &    0.12500 \\ \hline
    1.5 &   0.0 &   0.17106 &   0.18253 &   0.18453 &    0.18536
        &           0.18662 &   0.18663 (25)        &    0.18687 \\
    1.5 &   0.5 &   0.18283 &   0.18468 &   0.18553 &    0.18585
        &           0.18633 &                       &    0.18687 \\
    1.5 &   1.0 &   0.18515 &   0.18588 &   0.18620 &    0.18632
        &           0.18646 &                       &    0.18687 \\
    1.5 &   1.5 &   0.18684 &   0.18684 &   0.18687 &    0.18684
        &           0.18680 &                       &    0.18687 \\
    1.5 &   2.0 &   0.18878 &   0.18796 &   0.18759 &    0.18741
        &           0.18696 &                       &    0.18687 \\ \hline
    2.0 &   0.0 &   0.2076  &   0.2296  &   0.2321  &    0.2340
        &           0.2369  &   0.2392 (27)         &    0.2500  \\
    2.0 &   0.5 &   0.2283  &   0.2323  &   0.2342  &    0.2351
        &           0.2371  &                       &    0.2500  \\
    2.0 &   1.0 &   0.2325  &   0.2347  &   0.2358  &    0.2363
        &           0.2383  &                       &    0.2500  \\
    2.0 &   1.5 &   0.2357  &   0.2372  &   0.2379  &    0.2383
        &           0.2400  &                       &    0.2500  \\
    2.0 &   2.0 &   0.2402  &   0.2413  &   0.2417  &    0.2420
        &           0.2435  &                       &    0.2500  \\
 \end{tabular}
 \caption{Estimants $X(L_0,L_{\rm max})$ for the scaling dimension of the
   twist operator along with their extrapolations to the infinite-system
   limit. For $n_{\rm b}=1$ the value $X = 1/8$ was previously found to be
   independent of $n_{\rm g}$ [14].
   It is evident that this $n_{\rm g}$-independence holds for any value of
   $n_{\rm b}$, and in accordance herewith our final result is obtained by
   averaging the various extrapolants over $n_{\rm g}$.}
 \label{Tab:X}
\end{table}

\begin{multicols}{2}

Finally, the results for $x_2$ as obtained from parabolic fits for the gap
$f_0^{(2,0)}(L) - f_0^{(0,0)}(L)$ are shown in Table \ref{Tab:x2}. Again we
have $L_{\rm max} = 12$. Just like
in the case of $X$ we find the extrapolated values of $x_2$ to be
independent of $n_{\rm g}$, and final results are obtained by
averaging over this parameter.

\subsection{Entropy}
\label{sec:entropy}

Apart from the various universal quantities, such as the central charge
and the scaling dimensions, the transfer matrices also provide
numerical values for the residual entropy per vertex,
$s = f_0(\infty)$. In the limit $n_{\rg} \to 0$ of compact
polymers this quantity is of interest to the
protein folding community, due to the fact that native conformations
of all globular proteins are compact \cite{Chan-Dill89}.

Using our knowledge of the exact form of the finite-size
corrections of order $1/L^2$, \Eq{FSS-c}, 
we have obtained very accurate extrapolations to the limit
of an infinite system.\footnote{The logarithmic corrections to the
free energy implied by the ${\cal N}^{\gamma-1}$ term in
\Eq{thirumalai-scaling} does not pertain to the cylindrical geometry
implicit in our transfer matrix calculations. A similar remark applies
to the surface term $\kappa_{\rm s}^{{\cal N}^{(d-1)/d}}$.}
After subtracting the $1/L^2$ correction a
series of estimants $s(L,L_{\rm max})$ may be obtained by fitting the
residual size dependence to a pure $1/L^4$ form. 
The remaining $L$-dependence of these estimants turns out to be well
accounted for by a further $1/L^4$ fit, and in this way
we arrive at a final value for $s$. The error bar on the final value
can be estimated as its deviation from the most
accurate extrapolant, $s(L_{\rm max}-2,L_{\rm max})$.

The most accurate results are quite naturally found by employing this
procedure on $f_0^{(0,0)}(L)$, and they are shown in Table
\ref{Tab:entropy}. Results obtained by extrapolating the free energies
for other sectors of the transfer matrix containing strings are
consistent herewith but have error bars that are roughly 10 times
larger. If the fugacity of one of the strings equals two the error bars are
even larger, which is to be anticipated from the fact that
logarithmic corrections to the scaling dimensions are larger than
similar corrections to the central charge \cite{Cardy-log}.

\end{multicols}

\begin{table}
 \begin{tabular}{llllllrl}
      $n_{\rm b}$ &
      $n_{\rm g}$ & $x_1(4,12)$ & $x_1(6,12)$ & $x_1(8,12)$
        & Extrapolation         & Ref.~[14]                 & Exact   \\ \hline
    0.0 &   0.0   &    -0.2433  &    -0.2447  &    -0.2470
        &              -0.2500  &                           & -0.2500 \\ \hline
    0.5 &   0.0   &    -0.1328  &    -0.1295  &    -0.1303
        &              -0.1313  &                           & -0.1323 \\
    0.5 &   0.5   &    -0.01713 &    -0.01228 &    -0.01217
        &              -0.01217 &                           & -0.0131 \\ \hline
    1.0 &   0.0   &    -0.0440  &    -0.0423  &    -0.0430
        &              -0.0439  &    -0.0444 (1)            & -0.0446 \\
    1.0 &   0.5   &     0.0737  &     0.0763  &     0.0764
        &               0.0765  &     0.0750 (3)            &  0.0761 \\
    1.0 &   1.0   &     0.16608 &     0.16646 &     0.16657
        &               0.16663 &     0.1667 (1)            &  0.16667\\ \hline
    1.5 &   0.0   &     0.0267  &     0.0271  &     0.0264
        &               0.0255  &                           &  0.0260 \\
    1.5 &   0.5   &     0.1466  &     0.1472  &     0.1472
        &               0.1472  &                           &  0.1483 \\
    1.5 &   1.0   &     0.2411  &     0.2395  &     0.2395
        &               0.2394  &     0.242 (2)             &  0.2405 \\
    1.5 &   1.5   &     0.3196  &     0.3159  &     0.3156
        &               0.3156  &                           &  0.3162 \\ \hline
    2.0 &   0.0   &     0.0845  &     0.0848  &     0.0844
        &               0.0839  &                           &  0.1042 \\
    2.0 &   0.5   &     0.2070  &     0.2067  &     0.2071
        &               0.2076  &                           &  0.2295 \\
    2.0 &   1.0   &     0.3048  &     0.3021  &     0.3024
        &               0.3028  &     0.307 (2)             &  0.3250 \\
    2.0 &   1.5   &     0.3882  &     0.3841  &     0.3842
        &               0.3843  &                           &  0.4044 \\
    2.0 &   2.0   &     0.4640  &     0.4618  &     0.4635
        &               0.4657  &                           &  0.5000 \\
 \end{tabular}
 \caption{Scaling dimension $x_1$, corresponding to one string of each
         flavour.} 
 \label{Tab:x1}
\end{table}

\begin{multicols}{2}

In the special case of the equal-weighted six-vertex model,
$(n_{\rb},n_{\rg}) = (1,1)$, our value for $s$ is in excellent
agreement with the exact result due to Lieb \cite{Lieb67},
\begin{equation}
  s(1,1) = \frac32 \ln \left( \frac43 \right) \simeq 0.4315231 \cdots,
\end{equation}
and in the limit of two mutually excluding Hamiltonian walks,
$(n_{\rb},n_{\rg}) = (0,0)$, we are able to conjecture the result
\begin{equation}
  s(0,0) = \frac12 \ln (2) \simeq 0.3465735 \cdots.
\end{equation}
In fact, after having made this conjecture we discovered that the
numerical values of $f_0^{(1,1)}(L)$, i.e., the free energy per site 
in the  sector where we
enforce one string of each flavour, are {\em independent} of $L$
for $4 \le L \le 12$,  and equal to $\frac12 \ln(2)$ with full 16-digit
machine precision. Since the free energy per site in the thermodynamic
limit is  unchanged by the introduction of  a string
defect, this observation lends credibility to the correctness of the
above conjecture.

Our result in the compact polymer limit merits special
attention. Traditionally the entropy  is quoted in terms of the so-called
connective constant $\kappa = {\rm e}^{s(1,0)}$; see
\Eq{thirumalai-scaling}. Early approximations
due to Flory \cite{Flory42} and Huggins \cite{Huggins42} yielded
respectively
\begin{equation}
  \kappa_{\rm Flory} = \frac{z-1}{e} \simeq 1.104
\end{equation}
and
\begin{equation}
  \kappa_{\rm Huggins} = (z-1) \left(1 - \frac2z \right)^{z/2-1}
  = \frac32.
\end{equation}
Here $z=4$ is the coordination number of the square lattice. More
recently, $\kappa \simeq 1.472$ was found from transfer matrix calculations
\cite{Schmalz84} and $\kappa = 1.475 (15)$ by exhaustive computer
enumeration of short-chain configurations \cite{thirumalai}. Both
these results are very close to the mean-field value
$\kappa_{\rm MF} = \frac{z}{e} = 1.4715\cdots$ \cite{kappa_MF}, and it is
tempting to conclude that conformations of compact polymers are in fact 
described by mean-field theory \cite{thirumalai}. However, our result
\begin{equation}
\label{our_kappa}
  \kappa = 1.472801 (10)
\end{equation}
demonstrates that this is not the case.

\section{Discussion}
\label{sec_dis}

{}From the construction of the effective field theory of the
FPL${}^2$ model
some rather general conclusions regarding the scaling of compact polymers, and 
the relation between loop models and conformal field theory
can be drawn. It also  provides new  
insights into the three-state Potts antiferromagnet and the dimer loop 
model, which are identified with  specific points in 
the  phase diagram of the FPL${}^2$ model.   
We conclude the paper with a discussion of these topics.

\subsection{Compact polymers}

One of the  main motivations for studying fully
packed loop  models is  provided by  
compact polymers,   their scaling properties in particular. 
Just like polymers in the dilute and dense phase, compact
polymers form a critical geometrical system characterised by 
conformational exponents $\gamma$ and $\nu$. 
The  exponent $\gamma$ relates the number of conformations of the polymer to 
the number of monomers;  see Sec.~\ref{sec_com} for details. 
The other conformational  exponent ($\nu$) relates the linear
size of the polymer to the number of monomers. For compact structures 
it  has the trivial value  $1/2$ since these polymers are space filling. 

\end{multicols}

\begin{table}
 \begin{tabular}{rrrrrrlr}
    $n_{\rm b}$ &
    $n_{\rm g}$ & $x_2(4,12)$ & $x_2(6,12)$   & $x_2(8,12)$
        & Extrapolation       & Result        & Exact       \\ \hline
    0.0 &   0.0 &      0.0000 &        0.0000 &      0.0000
        &              0.0000 & 0.0000 (0)    &      0.0000 \\
    0.0 &   0.5 &      0.0000 &        0.0000 &      0.0000
        &              0.0000 &               &      0.0000 \\
    0.0 &   1.0 &      0.0000 &        0.0000 &      0.0000
        &              0.0000 &               &      0.0000 \\
    0.0 &   1.5 &      0.0000 &        0.0000 &      0.0000
        &              0.0000 &               &      0.0000 \\
    0.0 &   2.0 &      0.0000 &        0.0000 &      0.0000
        &              0.0000 &               &      0.0000 \\ \hline
    0.5 &   0.0 &      0.1279 &        0.1355 &      0.1372
        &              0.1389 & 0.1386 (2)    &      0.1386 \\
    0.5 &   0.5 &      0.1365 &        0.1371 &      0.1378
        &              0.1387 &               &      0.1386 \\
    0.5 &   1.0 &      0.1377 &        0.1374 &      0.1379
        &              0.1385 &               &      0.1386 \\
    0.5 &   1.5 &      0.1383 &        0.1375 &      0.1379
        &              0.1384 &               &      0.1386 \\
    0.5 &   2.0 &      0.1392 &        0.1376 &      0.1379
        &              0.1383 &               &      0.1386 \\ \hline
    1.0 &   0.0 &      0.2333 &        0.2447 &      0.2472
        &              0.2504 & 0.2495 (5)    &      0.2500 \\
    1.0 &   0.5 &      0.2488 &        0.2477 &      0.2484
        &              0.2493 &               &      0.2500 \\
    1.0 &   1.0 &      0.2514 &        0.2487 &      0.2490
        &              0.2494 &               &      0.2500 \\
    1.0 &   1.5 &      0.2538 &        0.2497 &      0.2495
        &              0.2492 &               &      0.2500 \\
    1.0 &   2.0 &      0.2573 &        0.2512 &      0.2504
        &              0.2494 &               &      0.2500 \\ \hline
    1.5 &   0.0 &      0.3197 &        0.3377 &      0.3416
        &              0.3466 & 0.3487 (26)   &      0.3506 \\
    1.5 &   0.5 &      0.3429 &        0.3425 &      0.3443
        &              0.3466 &               &      0.3506 \\
    1.5 &   1.0 &      0.3486 &        0.3457 &      0.3466
        &              0.3478 &               &      0.3506 \\
    1.5 &   1.5 &      0.3548 &        0.3497 &      0.3496
        &              0.3494 &               &      0.3506 \\
    1.5 &   2.0 &      0.3636 &        0.3561 &      0.3547
        &              0.3529 &               &      0.3506 \\ \hline
    2.0 &   0.0 &      0.3920 &        0.4202 &      0.4268
        &              0.4353 & 0.446 (12)    &      0.5000 \\
    2.0 &   0.5 &      0.4244 &        0.4277 &      0.4323
        &              0.4382 &               &      0.5000 \\
    2.0 &   1.0 &      0.4346 &        0.4348 &      0.4382
        &              0.4426 &               &      0.5000 \\
    2.0 &   1.5 &      0.4468 &        0.4452 &      0.4474
        &              0.4502 &               &      0.5000 \\
    2.0 &   2.0 &      0.4640 &        0.4618 &      0.4635
        &              0.4657 &               &      0.5000 \\
 \end{tabular}
 \caption{Scaling dimension $x_2$, corresponding to two black strings.}
 \label{Tab:x2}
\end{table}

\begin{multicols}{2}

Prior to our work,  
exact results have been obtained for compact polymers on the 
Manhattan \cite{Manhat} and the 
honeycomb \cite{Batch_FPL} lattice, and the mean-field value  
$\gamma=1$ was found in both cases. 
This value of $\gamma$ indicates that the two ends
of the compact polymer are independent at large distances. This follows from
the scaling relation 
$x_1=1-\gamma=0$, where the one-string dimension $x_1$ describes 
the probability $G_1(r)\sim r^{-2x_1}$ 
that the two chain ends are separated by a 
distance $r$. In this regard the scaling of compact polymers on the 
Manhattan and the honeycomb lattices is equivalent to that of ideal chains. 
Ideal chain configurations are described by simple  random walks 
for which each step is independent of the previous one.  

Here we have calculated the exact conformational exponent
$\gamma=117/112$ for compact polymers on the square lattice. The fact that 
$\gamma>1$ is tantamount to an effective {\em repulsion} between the
ends of the  
chain, indicating non-ideal behaviour.
Indeed, the fact that the connective constant $\kappa$ in \Eq{our_kappa} 
is larger than
its mean-field value indicates that the origin of this repulsion is
{\em entropic}. Earlier numerical studies of this 
problem utilising direct enumerations of chain conformations 
have failed to see any deviation from the 
ideal chain result $\gamma_{\rm MF}=1$ 
\cite{thirumalai}; we can attribute this 
to the fact that the 
actual difference  is indeed very small ($\gamma -\gamma_{\rm MF}=5/112$) 
and below the numerical accuracy previously achieved. The same comment 
can be made for the connective constant. 

Another interesting aspect of compact polymers is that their 
scaling properties are lattice dependent. This is in contrast to the 
dilute and dense case which are described by conformational exponents
that do not depend on the lattice type (e.g., honeycomb {\em versus} square). 
As remarked earlier this ``lack of universality''  
is due to a kind of  geometrical frustration 
that arises from  the fully packing constraint imposed on the loop models 
which are employed in studies of  compact polymers.  

Finally, the  field theory solution of the FPL${}^2$ model uncovered a
property of compact polymers that, to our knowledge, was not
previously anticipated.
The fact that there is a whole line of critical points in this loop model  
in the Hamiltonian walk limit ($n_{\rm b}\to 0$) indicates 
a continuum of universality classes described by compact polymers  
on the square lattice. In particular the exponent $\gamma$ can be 
changed continuously by adjusting the fugacity of the loops uncovered 
by the polymer. The loop weight of the uncovered (grey) loops can be thought 
of as an effective interaction amongst the monomers, albeit a non-local 
one. A similar effect of interactions on directed self-avoiding walks 
was discovered by Cardy \cite{cardy_gam} from a field theoretical analysis 
of the problem. The existance of a continuously varying $\gamma$ in this 
case was recently challanged by numerical results \cite{trovato_gam}.   

\end{multicols}

\begin{table}
 \begin{tabular}{rrrrrrrrl}
  $n_{\rm b}$ & $n_{\rm g}$ &
  $f_0^{(0,0)}(4)$  & $f_0^{(0,0)}(6)$  & $f_0^{(0,0)}(8)$  &
  $f_0^{(0,0)}(10)$ & $f_0^{(0,0)}(12)$ & $f_0^{(0,0)}(14)$ & s \\ \hline
   0.0 & 0.0 &
         0.17328680 &        0.28881133 &        0.31784496 &
         0.32923359 &        0.33490107 &        0.33815371 & 0.346575 (14)
         \\ \hline
   0.5 & 0.0 &
         0.26740000 &        0.33317928 &        0.35057672 &
         0.35745438 &        0.36088114 &        0.36284872 & 0.367950 (9) \\
   0.5 & 0.5 &
         0.35063553 &        0.37668215 &        0.38371283 &
         0.38639599 &        0.38769210 &        0.38842126 & 0.390258 (3)
         \\ \hline
   1.0 & 0.0 &
         0.32923947 &        0.36764369 &        0.37752555 &
         0.38137032 &        0.38327066 &        0.38435762 & 0.387166 (7) \\
   1.0 & 0.5 &
         0.40772622 &        0.41103439 &        0.41126990 &
         0.41111188 &        0.41095017 &        0.41082815 & 0.410405 (2) \\
   1.0 & 1.0 &
         0.46298939 &        0.44576535 &        0.43960110 &
         0.43671524 &        0.43513763 &        0.43418273 & 0.4315233 (4)
         \\ \hline
   1.5 & 0.0 &
         0.37601935 &        0.39599984 &        0.40063320 &
         0.40233073 &        0.40314475 &        0.40360330 & 0.404771 (5) \\
   1.5 & 0.5 &
         0.45180855 &        0.43964968 &        0.43509788 &
         0.43291844 &        0.43171625 &        0.43098591 & 0.4289459 (10) \\
   1.5 & 1.0 &
         0.50624745 &        0.47501911 &        0.46431698 &
         0.45948057 &        0.45688890 &        0.45533728 & 0.4510742 (17) \\
   1.5 & 1.5 &
         0.54930614 &        0.50513652 &        0.49006459 &
         0.48331974 &        0.47972832 &        0.47758588 & 0.471726 (2)
         \\ \hline
   2.0 & 0.0 &
         0.41389271 &        0.42018005 &        0.42097629 &
         0.42111147 &        0.42113891 &        0.42114428 & 0.421145 (6) \\
   2.0 & 0.5 &
         0.48795109 &        0.46429984 &        0.45622604 &
         0.45257417 &        0.45061901 &        0.44945033 & 0.4462607 (10) \\
   2.0 & 1.0 &
         0.54202495 &        0.50046092 &        0.48641918 &
         0.48016010 &        0.47683419 &        0.47485271 & 0.4694505 (18) \\
   2.0 & 1.5 &
         0.58515036 &        0.53158535 &        0.51333087 &
         0.50520022 &        0.50088581 &        0.49831761 & 0.491323 (3) \\
   2.0 & 2.0 &
         0.62122666 &        0.55918707 &        0.53795845 &
         0.52850379 &        0.52348906 &        0.52050483 & 0.5123870 (19) \\
 \end{tabular}
 \caption{Residual entropy $s$, obtained by extrapolating
          $f_0^{(0,0)}(L)$ to the infinite-system limit.}
 \label{Tab:entropy}
\end{table}

\begin{multicols}{2}

\subsection{Relation to other models} 

The FPL${}^2$ model is a loop model which exhibits a 
two-dimensional manifold of fixed points in its phase diagram.  
Certain points in the critical region map to previously studied lattice 
models and here we comment on the relevance of our results for these models. 

\subsubsection{Dimer loop model}

The dimer loop model studied by Raghavan {\em et al.}~\cite{raghavan} is the
$n_{\rb}=2$, $n_{\rg}=1$ FPL${}^2$ model; see \Fig{Fig:phasediag}. 
The dimer loop model is defined by
placing black and
white dimers on the square lattice so that every vertex of the lattice is
covered by exactly one black and one white dimer. Every such configuration
is given equal weight. The mapping to the FPL${}^2$ model is achieved by 
identifying the bonds covered by dimers as making up the black loops, whilst 
the uncovered bonds form the grey loops.  
The original motivation 
for studying this dimer problem is that it leads to a height model 
with a two-component height; cfr.~the traditional dimer model which is
described by a single component height. 

Performing Monte Carlo simulations of the dimer loop model
Raghavan {\em et al.}~reached the conclusion that 
one of the two height components is rough whilst the other one is 
``anomalously smooth'', i.e., its structure function decays at 
small wave-vectors ${\bf q}$ slower then $1/{\bf q}^2$; a $1/{\bf q}^2$ 
dependence is to be expected in a  Gaussian field theory.

In light of our results we would conclude that the dimer loop model 
is critical with a central charge $c=2$.  This follows from 
\Eq{c_charge1} for $n_{\rb}=2$ and $n_{\rg}=1$. The two components of the 
height found by Raghavan {\em et al.}~should therefore both be rough, each 
contributing one to the central charge ($c=1+1$). Furthermore, 
we believe that the observed 
anomalous behaviour of one of the heights can be attributed  
to the fact that this model is exactly 
at the boundary of the critical region of the FPL${}^2$ model. We observe a 
similar effect  
in our numerical transfer matrix results which show largest deviations 
from the proposed exact formulae for loop fugacities at the critical-region 
boundary. The culprit might be logarithmic corrections due to the presence of
marginal operators.    
To check this 
hypothesis and reconcile it with the fact that no such effects are seen 
in Monte Carlo  simulations of the four-colouring model \cite{jk_prb}   
($n_{\rb}=n_{\rg}=2$), which is also at the boundary of the critical region,  
simulations of the dimer-loop model for larger system sizes would be welcome.

\subsubsection{Three-state Potts antiferromagnet}

The critical ground state of three-state Potts antiferromagnet 
maps to the equal-weighted six-vertex
model \cite{widom} which is the $n_{\rb}=n_{\rg}=1$ point in the 
critical region of the FPL${}^2$ model; see \Fig{Fig:phasediag}.  
Along the line
$n_{\rb}=n_{\rg}$ the colouring representation of the FPL${}^2$ model
has the additional symmetry with respect to cyclic permutations of the
four colours; see Sec.~\ref{loop_ansatz}.
This explains the origins of the $\Zs_4$ symmetry found by 
Saleur for the {\em three}-state  Potts antiferromagnet \cite{saleur_potts}.

\subsubsection{Folding model}

The folding model of the square-diagonal lattice
recently investigated by Di Francesco \cite{DiFrancesco} maps onto a
constrained version of the $(n_{\rb},n_{\rg})=(2,2)$ \F2 model.
The constraint consists in allowing only the vertices 1, 3, 5 and 6 of
Fig.~\ref{Fig:vertices} for sites on the even sublattice, and similarly
vertices 2, 4, 5 and 6 on the odd sublattice.

We have modified our transfer matrices to take this constraint into
account. Our result for the folding entropy, $s = 0.4604(4)$, is in
complete agreement with Ref.~\cite{DiFrancesco}.\footnote{Our
normalisation is ``per vertex'' whilst that of Di Francesco is ``per
triangle''. Accordingly we find twice his result.} Interestingly enough the
finite-size scaling of the gaps in the eigenvalue spectrum seems to
indicate that the model is not critical for general values of the loop
fugacities. From the field theory of the \F2 model we should be able to 
to understand why the constraint imposed by the folding  model 
leads to a relevant perturbation which takes  the system away from
criticality. This we leave as an interesting open question. 
Incidentally, the situation is very reminiscent of the reformulation
of the $Q$-state Potts model in terms of a staggered vertex model.
Only at the critical point are the vertex weights on the even and odd
sublattices identical, thus allowing for an exact solution of the
model \cite{Baxter}.

\subsection{Conformal field theory} 

The Liouville field theory proposed for the effective theory of the
FPL${}^2$ model
in the critical region is conformally invariant. Each point in the critical 
phase is  characterised by the 
central charge and the scaling dimensions of primary fields, which are 
associated with  electric and magnetic charges in the Coulomb gas. 
For generic values of the 
loop fugacities the background charge ${\bf e}_0$ is not commensurate with 
the electric charges that make up the lattice ${\cal R}^*$. This implies that 
amongst the electric  operators  
there will be many (an infinite number, in fact) that 
have negative dimensions, signaling the non-unitary nature of the 
conformal field theory. Non-unitary CFT's
appear in many other critical geometrical models, critical percolation being 
the best known example.

Liouville field theory provides the Euclidean action for the 
Coulomb gas description of conformal field theories proposed by Dotsenko and
Fateev \cite{dots_fateev}. As such it contains the so-called screening
charges which are the 
vertex operators that make up the Liouville potential. In the original 
formulation these charges were introduced on formal grounds so as to 
ensure the existence of non-vanishing four-point correlation 
functions in the theory. 
In order for these vertex operators not to disrupt the 
conformal symmetry of the 
modified Gaussian model (the modification is the addition of the boundary 
term to the gradient-square action) they are necessarily marginal, i.e., 
their scaling dimension is 2. 

Here we have found a physical interpretation of the screening charges. Their
r\^{o}le in loop models is to ensure that the number of large 
loops from scale to 
scale stays of order one; this translates into the statement that the loop 
fugacities do not flow under the action of the renormalisation group. 

The fact that we have a concrete physical interpretation of the screening
charges directly leads to the calculation of the elastic constants in the 
Liouville field theory. In the traditional Coulomb gas approach these 
coupling constants are calculated by comparing to formulae derived from 
an exact solution of the model. Once these constants are known marginal 
vertex operators that play the r\^{o}le of screening charges can be written 
down. Our construction basically  reverses this procedure, and by doing so 
{\em makes no reference to an exact solution}.  

Finally, we end with a speculatory 
note concerning the prospects of solving the 
FPL${}^2$ model via Bethe Ansatz. Namely, 
all loop models to date have been solved by this method after  
mapping them to a vertex model, following a procedure 
analogous to the one outlined in Sec.~\ref{sec_height}. 
This does not seem to work 
for the FPL${}^2$ model, at least not along the $n_{\rg}=1$ line
\cite{Batchelor96}. Why this is so is an interesting open question. 

One possibility is that the {\em full} FPL${}^2$ model needs to be
considered as  
opposed to the FPL model studied by Batchelor {\em et al.}~for which 
$n_{\rg}=1$ is fixed. A more intriguing possibility is that 
a Bethe Ansatz solution might be hindered (or made more difficult) 
by the non-trivial 
elasticity displayed by the FPL${}^2$ model in its interface representation. 
This statement 
we base solely on the observation that all previously solved loop 
models are simple  as interface models in the sense that the height
fluctuations are described by a single elastic constant. For the 
FPL${}^2$ model, as described in Sec.~\ref{sec_LFT}, 
the stiffness tensor consists of 
three independent components. Whether indeed the interface representation of
the loop model has any bearing on its Bethe Ansatz solvability remains to 
be seen.

\begin{acknowledgements}

Useful discussions with M. Aizenman, J.~L.~Cardy, B. Duplantier, D.S. Fisher, 
C.L.~Henley, T.~Hwa, T.~Prellberg, T.~Spencer, F.~Y.~Wu, and C. Zeng
are gratefully acknowledged.
The authors would like to thank the Institute for Theoretical Physics at Santa
Barbara, where this collaboration was initiated, for its warm hospitality.
This research was supported in part by the Engineering and Physical Sciences
Research Council under Grant GR/J78327, and by the National Science Foundation
under Grant PHY94-07194, and DMS 9304580. 

\end{acknowledgements}

\appendix
\section{Dimensions of electric and magnetic operators}
\label{app:Gauss}

We calculate the  scaling dimensions of electric and magnetic 
operators in the Coulomb gas theory described by the action 
\be{G_action}
S_{\rm CG} = \frac{1}{2} \int \! d^2{\bf x} \ g_{\alpha} (\bp H^{\alpha})^2
             +  \frac{{\rm i}}{4 \pi} \int \! d^2{\bf x} \ ({\bf E}_0 
             \cdot {\bf H}) {\cal R} \ ,  
\ee
where ${\cal R}$ is the scalar curvature. We are interested in the 
situation when the height field is defined on a flat surface, in which case
${\cal R}$ is zero everywhere except at the boundaries.

\subsection{Electric charges}

The scaling dimension $x({\bf E})$,  of the electric-type  operator 
$\exp({\rm i}{\bf E}\cdot{\bf H}({\bf x}))$,  
follows from the two-point  function 
\be{cor_el} 
\left< {\rm e}^{i {\bf E} \cdot {\bf H}({\bf x})} 
      {\rm e}^{-i ({\bf E}-2{\bf E}_0) \cdot {\bf H}({\bf y})} \right> \sim 
       |{\bf x} - {\bf y}|^{-2 x({\bf E})},
\ee
where the expectation value is with respect to the measure defined by 
the action $S_{\rm CG}$. The extra electric charge $2{\bf E}_0$ appears
due to the charged boundary conditions enforced by the curvature term 
in the Coulomb gas action, \Eq{G_action}.

We break up the calculation into two parts. 
First we  calculate the two-point function, \Eq{cor_el}, 
in the absence of the background charge (${\bf E}_0=0$). 
We make use of the property of Gaussian integrals, 
\bea{gaus_cor_el} 
  \lefteqn{\left< {\rm e}^{i {\bf E} \cdot {\bf H}({\bf x})}
       {\rm e}^{-i{\bf E} \cdot {\bf H}({\bf y})} \right> =} \\  
  & & \ \ \ \
  \exp\left( - \frac{1}{2}(E_\alpha)^2 \left< (H^\alpha({\bf x})
             -H^\alpha({\bf y}))^2 \right> \right), \nonumber
\eea
and of the known propagator for the massless scalar field in two dimensions
(where we have dropped the regulators at large and small distances),
\be{gaus_prop}
 \left< (H^\alpha({\bf x})-H^\alpha({\bf y}))^2 \right> = \frac{1}{\pi 
 g_\alpha}  \ln|{\bf x} - {\bf y}| \ .
\ee
Combining the above two equations  and comparing the result to \Eq{cor_el},
we find 
\be{E-dim-0}
2 x^{(0)}_{\rm e}({\bf E}) = \frac{1}{2 \pi g_\alpha} (E_\alpha)^2  \ ;
\ee   
the superscript $(0)$ is there to remind us that this formula is valid
only for ${\bf E}_0=0$. 

This result for the two-point function  can be rewritten as
\be{gaus_cor2}
      \left< {\rm e}^{i {\bf E} \cdot {\bf H}({\bf x})}
             {\rm e}^{-i{\bf E} \cdot {\bf H}({\bf y})} \right> = 
      \exp[{\cal E}^{(0)}_{\bf E}({\bf x},{\bf y})],
\ee
where 
\be{el_energy}
{\cal E}^{(0)}_{\bf E}({\bf x},{\bf y})= - 
\frac{1}{2 \pi g_\alpha} (E_\alpha)^2 \ln|{\bf x}-{\bf y}|
\ee
is the energy for two (vector) electric charges interacting via the 
two-dimensional Coulomb force; in this language $S_{\rm E}$ is
the energy of the electrostatic field set up by the electric charges
$\pm{\bf E}$, 
expressed in terms of the electrostatic potential ${\bf h}$.   
This seemingly trivial rewriting makes the calculation 
of $x({\bf E})$, the electric dimension in the presence of a background 
charge, physically transparent.

To properly take into account the curvature term we 
define  the height field over  a 
disc of radius $R$, instead of the infinite plane, keeping in mind that 
at the end of the calculation we need to take the  limit $R\to \infty$. 
In the case of the disc ${\cal R} = 8\pi\delta(R)$, and the curvature term 
introduces a charge 
$-2{\bf E}_0$ at the disc boundary. Therefore, the vacuum 
of the modified Coulomb gas must contain a {\em floating charge} 
$+2{\bf E}_0$ in the disc interior, and the electrostatic energy of this 
charged vacuum is ${\cal E}^{(0)}_{2 {\bf E}_0}(0, R)=-4 {\rm E}_{0\alpha}^2 
\ln(R)/2\pi g_{\alpha}$. Now, to find the scaling 
dimension of a vertex operator of charge ${\bf E}$, we imagine placing 
charges $+{\bf E}$ and $-{\bf E}$ at points ${\bf x}$ and ${\bf y}$ in the 
disc interior, and we calculate the total electrostatic energy with respect to 
the charged vacuum. The floating 
charge being positive will coalesce with the negative  charge  $-{\bf E}$. 
Using Coulombs law, \Eq{el_energy}, we then calculate the interaction 
energy of 
charges $+{\bf E}$ at {\bf x}, $-{\bf E}+2{\bf E}_0$ at ${\bf y}$, and 
$-2{\bf E}_0$ at $R$, keeping in mind $R\gg|{\bf x}-{\bf y}|$. 
The final result 
\be{el_energy2}
  {\cal E}_{{\bf E}}({\bf x}, {\bf y}) =
    - \frac{1}{2 \pi g_\alpha} E_\alpha (E_\alpha - 2E_{0\alpha}) 
    \ln|{\bf x}-{\bf y}| 
\ee
is obtained after the energy of the charged vacuum is subtracted. 
Now it is a simple 
matter to read off the scaling dimension as the negative coefficient in front 
of the logarithm,
\be{E-dim-1}
 2 x({\bf E}) = \frac{1}{2 \pi g_\alpha} E_\alpha 
 (E_\alpha - 2 E_{0\alpha}) \ . 
\ee   
This result can be derived in a more rigorous fashion by constructing
the stress-energy tensor for the field theory $S_{\rm CG}$ and calculating 
its  operator product  with the vertex operator 
$\exp(i{\bf E}\cdot{\bf H})$ \cite{dots_fateev}.

\subsection{Magnetic charge}

To calculate  the magnetic dimension $x({\bf M})$ we consider 
the ratio of partition functions, 
\be{cor_m} 
Z_{>{\bf M}}({\bf x}, {\bf y}) / Z_> \sim |{\bf x} - {\bf y}|^{-2 
x({\bf M})} \ .
\ee
$Z_{>{\bf M}}({\bf x},{\bf y})$ is the sum (path integral) over height 
configurations where a vortex and an antivortex, of 
topological charge $\pm{\bf M}$, are placed at positions 
${\bf x}$ and ${\bf y}$ of the basal plane, whilst $Z_>$ is the unconstrained 
sum:
\be{Z>_sum}
 Z_> = \int \! {\cal D}{\bf H} \ \exp\left( -\frac{1}{2} \int \! d^2{\bf x} 
  \ g_{\alpha} (\bp H^{\alpha})^2 \right) \ . 
\ee
Here we have dropped the curvature term since it does not affect  
correlation functions of  magnetic operators. 

We can use the electrostatic analogy once again. Namely,
we consider the interaction energy between two topological defects,  
${\cal E}_{\bf M}({\bf x},{\bf y})$. Since $Z_>$  
is a Gaussian path integral, it follows that 
\be{cor_m_EN}
Z_{>{\bf M}}({\bf x}, {\bf y}) / Z_> = 
\exp[{\cal E}_{\bf M}({\bf x},{\bf y})] \ ,
\ee
where 
\be{mag_en}
- {\cal E}_{\bf M}({\bf x},{\bf y}) = 
   \frac{g_\alpha}{2 \pi} (M^\alpha)^2  \ln|{\bf x} - {\bf y}| \ . 
\ee
The above interaction energy is calculated as the Gaussian action of the
the  classical configuration of the height field, ${\bf h}_{\rm c}$.   
${\bf h}_{\rm c}$ 
solves the classical equations of motion (Laplace's equation) with   
boundary conditions dictated by the presence of topological defects at 
${\bf x}$ and ${\bf y}$ \cite{chaikin-lubensky}. 
The scaling dimension of a magnetic-type operator is then the
coefficient in front of the logarithm in \Eq{mag_en}, 
\be{mag_dim_fin}
2 x({\bf M}) = \frac{g_\alpha}{2 \pi} (M^\alpha)^2  \ .
\ee

\section{Enumeration of the connectivities}
\label{app:Connect}

The implementation of the transfer matrix (TM) for the FPL$^2$ model on a
cylinder of width $L$ and length $M$ requires an enumeration of the
possible connectivity states of the $L$ points on the dangling edges of row
$M$. Each of these $L$ points can either
\begin{enumerate}
 \item be connected by ${\cal G}_M$ to one of the dangling edges of row 0
   through a string of flavour $i = {\rm b},{\rm g}$, or
 \item be connected by ${\cal G}_M$ to one and only one other point in row
   $M$ through a loop segment of flavour $i = {\rm b},{\rm g}$.
\end{enumerate}
A suitable representation of this information is furnished by a double
state vector
\begin{equation}
  \left( \begin{array}{c}
           i^{\rm b}_1 i^{\rm b}_2 i^{\rm b}_3 \ldots i^{\rm b}_L \\
           i^{\rm g}_1 i^{\rm g}_2 i^{\rm g}_3 \ldots i^{\rm g}_L \\
         \end{array}
  \right),
  \label{state}
\end{equation}
which we shall refer to as the {\em index representation}. The indices
$i^{\rm b}_k$ ($k=1,2,\ldots,L$) are defined as follows:
\begin{enumerate}
 \item $i^{\rm b}_k = i^{\rm b}_l$ is a (non-unique) positive integer
   if and only if points $k$ and $l$ are interconnected through a
   black string.
 \item $i^{\rm b}_k = 0$ if and only if point $k$ touches a grey string or
   loop segment.
 \item $i^{\rm b}_k = -1$ if and only if point $k$ is connected to a dangling
   edge of row 0 through a black string.
\end{enumerate}
A similar definition is true for the indices $i^{\rm g}_k$ provided that one
reads ``grey'' instead of ``black'' and {\em vice versa}. Two index
representations are said to be identical if they are so up to the
arbitrariness of the choice of positive integers. Also note that if 
$i^{\rm b}_k \neq 0$ we have $i^{\rm g}_k = 0$ and conversely.

A restriction on those indices that take positive values
follows from the fact that loops of the same flavour are not allowed to
intersect. Namely, if $j < k < l < m$ the equalities
$i^{\rm b}_j = i^{\rm b}_l$ and $i^{\rm b}_k = i^{\rm b}_m$ cannot
both be true. So in addition to being pairwise these connectivities
are also {\em well-nested} \cite{Blote82}. The same is true 
for the grey indices, whereas there are no such restrictions when both
flavours are involved. Indeed, connectivity states with
$i^{\rm b}_j = i^{\rm b}_l$ and $i^{\rm g}_k = i^{\rm g}_m$ are
explicitly allowed by the last two vertices shown in Fig.~\ref{Fig:vertices}.

In practice we are only interested in the first few eigenvalues of TMs
having a definite number of strings of each flavour. The relevant sectors
of the TM are denoted ${\bf T}^{(s_{\rm b},s_{\rm g})}$, where $s_i$
is the number of strings of flavour $i = {\rm b},{\rm g}$. The fully
packing constraint means that we can only examine system sizes $L$
that have the same parity as $s_{\rm b} + s_{\rm g}$. The various
sectors have different physical interpretations and 
each requires a different enumeration of the connectivity states. Since the
two flavours enter at an equal footing in the partition function,
Eq.~(\ref{partition}), we only need consider $s_{\rm b} \ge s_{\rm g}$. The
${\bf T}^{(0,0)}$ sector contains information about the free energy and
the energy-like correlation length.
The geometrical scaling dimensions $x_1$ and $x_2$ can be obtained
from the ${\bf T}^{(1,1)}$ and the ${\bf T}^{(2,0)}$ sectors respectively.
Finally the sector ${\bf T}^{(1,0)}$ gives the scaling dimension of the
twist-like operator.

Whilst the index representation contains all information necessary for
determining the value of a given entry in the TM it is obviously not
suitable for labeling the entries. We therefore need another
representation, the so-called {\em number representation}, in which the
connectivities are labeled by the integers
$1,2,\ldots,C_L^{(s_{\rm b},s_{\rm g})}$, where $C_L^{(s_{\rm b},s_{\rm g})}$
is the number of different connectivity states in
the relevant sector. The practical implementation of the TMs relies on
the mapping from the index to the number representation and its inverse.

We shall now consider, one by one,  the various sectors of the TM.

\subsection{${\bf T}^{(0,0)}$ sector}

When no strings are present all the $L$ dangling edges of row $M$ are
pairwise connected with either a black or a grey loop segment. In
particular $L$ must be even. For any particular connectivity we can then
decompose $L$ as $L = 2 p_{\rm b} + 2 p_{\rm g}$, where $p_i \ge 0$
is the number of 
{\em pairs} of dangling edges covered by a flavour $i$ loop segment.
Since loops of different flavours are allowed to cross
(see Fig.~\ref{Fig:vertices}) the total number of connectivities is
\begin{equation}
  C_L^{(0,0)} = \sum_{L = 2 p_{\rm b} + 2 p_{\rm g}}
                {L \choose 2 p_{\rm b}} c_{p_{\rm b}} c_{p_{\rm g}},
  \label{C_L(0,0)}
\end{equation}
where $c_p$ is the number of pairwise well-nested connectivities of $2p$
points. The $c_p$'s were first considered in the context of the Potts model
\cite{Blote82}, but were also found to play a central r\^{o}le in the TM
formulation of the O($n$) model \cite{Blote89}. We shall now briefly recall
how they are evaluated.

Consider a well-nested pairwise connectivity of $2p$ points given by the
index representation $(i_1 i_2 \ldots i_{2p})$. A recursion relation
follows from observing that $i_1 = i_{2k}$ for precisely one integer
$k \ge 1$. According to the well-nestedness criterion the sub-sequences
$(i_2 i_3 \ldots i_{2k-1})$ and $(i_{2k+1} i_{2k+2} \ldots i_{2p})$ are
both well-nested, and indices occurring in one of them do not occur in the
other. Hence for $p \ge 1$
\begin{equation}
  c_p = \sum_{k=1}^p c_{k-1} c_{p-k},
\end{equation}
and $c_0 = 1$. By means of the generating function
$P(x) = \sum_{p=0}^{\infty} c_p x^p$ it is readily shown \cite{Blote89}
that
\begin{equation}
  c_p = \frac{(2p)!}{p!(p+1)!},
  \label{c_p}
\end{equation}
and that asymptotically $c_p \sim 4^p$.

Using Eqs.~(\ref{C_L(0,0)}) and (\ref{c_p}) we can now compute explicit
values for the $C_L^{(0,0)}$. These are shown for $2 \le L \le 16$ in Table
\ref{Tab:connect}.

\end{multicols}

\begin{table}
 \begin{tabular}{rrrrr|rr}
 $L$ & $4^L$ & $C_L^{(0,0)}$ & $C_L^{(1,1)}$ & $\tilde{C}_L^{(2,0)}$
     & $L$   & $C_L^{(1,0)}$                           \\ \hline
   2 &            16 &         2 &           2 &          1 &  1 &         1 \\
   4 &           256 &        10 &          24 &         12 &  3 &         6 \\
   6 &         4,096 &        70 &         300 &        150 &  5 &        50 \\
   8 &        65,536 &       588 &       3,920 &      1,960 &  7 &       490 \\
  10 &     1,048,576 &     5,544 &      52,920 &     26,460 &  9 &     5,292 \\
  12 &    16,777,216 &    56,628 &     731,808 &    365,904 & 11 &    60,984 \\
  14 &   268,435,456 &   613,470 &  10,306,296 &  5,153,148 & 13 &   736,164 \\
  16 & 4,294,967,296 & 6,952,660 & 147,232,800 & 73,616,400 & 15 & 9,202,050 \\
 \end{tabular}
 \caption{The number $C_L^{(s_{\rm b},s_{\rm g})}$ of FPL$^2$
   connectivity states for $L$ dangling edges accommodating $s_i$
   strings of flavour $i = {\rm b},{\rm g}$. Only values of $L$ with
   the same parity as $s_{\rm b} + s_{\rm g}$ are shown. When more
   than one string of any flavour is present further restrictions than the
   well-nestedness criterion apply, as described in the text. Accordingly
   the number $\tilde{C}_L^{(2,0)}$ is merely a useful upper limit on the
   true $C_L^{(2,0)}$.
   The efficiency of writing the TMs in the connectivity basis can be
   appreciated by comparing  $C_L^{(0,0)}$ 
   to $4^L$, the latter being the dimensions of the TM
   written in the conventional colour basis, where every dangling end is
   labeled independently by ${\bf A}$, ${\bf B}$, ${\bf C}$ or ${\bf D}$.}
 \label{Tab:connect}
\end{table}

\begin{multicols}{2}

For obvious reasons we shall call the function
\begin{equation}
  \rho(i_1 i_2 \ldots i_{2p}) = k
\end{equation}
defined by $i_1 = i_{2k}$ the {\em cut function} of the index
representation $(i_1 i_2 \ldots i_{2p})$. A complete ordering of the
well-nested sequences is now induced by applying the cut function first to
the whole sequence, then recursively to its right and finally to its left
part \cite{Blote82,Blote89}. Accordingly, the mapping from the index to the
number  representation for a well-nested one-flavour connectivity is
accomplished by
\begin{equation}
  \sigma(i_1 i_2 \ldots i_{2p}) = \left \{ \begin{array}{l}
    1 \mbox{ if } p \le 1 \\
    \begin{array}{l}
      \sum_{l=1}^{k-1} c_{l-1} c_{p-l} + \sigma(i_2 \ldots i_{2k-1}) \\
      + [\sigma(i_{2k+1} \ldots i_{2p}) - 1] c_{k-1} \mbox{ otherwise,}
    \end{array}
    \end{array} \right.
  \label{sigma}
\end{equation}
where the $c_p$ are given by \Eq{c_p}.

To give a complete specification of the connectivity of any {\em one}
flavour in the state (\ref{state}) we need to keep track of the positions
of those indices that are zero. For a fixed number of $z$ zero indices this
is accomplished by the lexicographic ordering
\begin{equation}
  \psi(i_1 i_2 \ldots i_L) = \left \{ \begin{array}{l}
    1 \mbox{ if } L=1 \mbox{ or } z=L \\
    \psi(i_2 i_3 \ldots i_L) \mbox{ if } i_1 \neq 0 \\
    {L-1 \choose z} + \psi(i_2 i_3 \ldots i_L) \mbox{ if } i_1 = 0,
  \end{array} \right.
\end{equation}
assigning the lowest value to the sequence with all the zeros accumulated
to the right.

The number representation of the two-flavour state (\ref{state}) is
now obtained by first ordering according to the number of indices
$i^{\rm b}_k$ being zero, then lexicographically ordering the positions of these
zero indices, and finally using the ordering (\ref{sigma}), first on the
well-nested subsequence of non-zero black indices and then on the
corresponding grey subsequence. More precisely, the mapping from the
index to the number representation in the $(s_{\rm b},s_{\rm g}) = (0,0)$
sector is given by 
\begin{eqnarray}
  \lefteqn{ \phi^{(0,0)}
  \left( \begin{array}{c} {\bf i}^{\rm b} \\ {\bf i}^{\rm g}
      \\ \end{array} \right)
  = \sum_{k=p_{\rm b}+1}^{L/2} {L \choose 2k} c_k c_{L/2-k} } \\
  \label{phi}
  & & \ \ \ \
    + [\psi({\bf i}^{\rm b}) - 1] c_{p_{\rm b}} c_{p_{\rm g}} + 
      [\sigma({\bf \tilde{i}}^{\rm b}) - 1] c_{p_{\rm b}} +
       \sigma({\bf \tilde{i}}^{\rm g}),
      \nonumber
\end{eqnarray}
where ${\bf i}^{\rm b} = (i^{\rm b}_1 i^{\rm b}_2 \ldots i^{\rm b}_L)$
denotes the sequence of black indices and ${\bf \tilde{i}}^{\rm b}$
the subsequence of the $p_{\rm b}$ pairs of non-zero indices (and, of
course, similarly for the grey flavour).

The inversion of \Eq{phi}, so as to furnish a mapping from the
number to the index representation, is straightforward if we know how to
invert the functions $\sigma$ and $\psi$. Details on this have already been
given in Ref.~\cite{Jesper4}.

\subsection{${\bf T}^{(1,0)}$ sector}

In the case of one black string spanning the length of the cylinder
the number of dangling edges in row $M$ can be written as
$L = 2p_{\rm b} + 2p_{\rm g} + 1$, where the $p_i$ have the same
meaning as above. In particular $L$ must be odd.

The presence of {\em one} string of either flavour does not impose any
additional restrictions on the connectivity states of the subsequence of
positive indices of that flavour. Indeed, if the position of the string is
given by $i^{\rm b}_r = -1$ the non-zero subsequence of
$(i^{\rm b}_{r+1} \ldots i^{\rm b}_L i^{\rm b}_1 \ldots i^{\rm b}_{r-1})$
is still well-nested, and the arguments given above apply. The number
of connectivity states is therefore found by multiplying the $L$
possible positions of the string by the number of
$(s_{\rm b},s_{\rm g}) = (0,0)$ states of the remaining $L-1$ points
\begin{equation}
  C_L^{(1,0)} = L C_{L-1}^{(0,0)}.
  \label{C_L(1,0)}
\end{equation}
Explicit values are shown in Table \ref{Tab:connect}.

Similarly the mapping from the index to the number representation is found
by first ordering after the position $r$ of the string, and then after the
value of $\phi^{(0,0)}$ taken of the remaining indices
\begin{eqnarray}
  \phi^{(1,0)} \left( \begin{array}{c}
                        i^{\rm b}_1 i^{\rm b}_2 \ldots i^{\rm b}_L \\
                        i^{\rm g}_1 i^{\rm g}_2 \ldots i^{\rm g}_L \\
                      \end{array}
               \right) &=& (r-1) C_{L-1}^{(0,0)} \\
                       &+&
  \phi^{(0,0)} \left( \begin{array}{c}
                        i^{\rm b}_1 \ldots i^{\rm b}_{r-1}
                        i^{\rm b}_{r+1} \ldots i^{\rm b}_L \\
                        i^{\rm g}_1 \ldots i^{\rm g}_{r-1}
                        i^{\rm g}_{r+1} \ldots i^{\rm g}_L \\
                       \end{array}
               \right). \nonumber
\end{eqnarray}

\subsection{${\bf T}^{(1,1)}$ sector}

When one string of each flavour is present $L = 2p_{\rm b} + 2p_{\rm g} + 2$
must be even, and again it suffices to augment the considerations from the
${\bf T}^{(0,0)}$ case by some book-keeping as to the positions of the two
strings. Explicit values of
\begin{equation}
  C_L^{(1,1)} = L(L-1)C_{L-2}^{(0,0)}.
  \label{C_L(1,1)}
\end{equation}
are shown in Table \ref{Tab:connect}.

Letting $r_i$ denote the position of the string of flavour
$i = {\rm b},{\rm g}$ we find that
\begin{eqnarray}
  &\phi^{(2,0)}& \left( \begin{array}{c} {\bf i}^{\rm b} \\ {\bf i}^{\rm g}
                     \\ \end{array} \right)
  = [(r_{\rm b}-1)(L-1) + (|r_{\rm g}-r_{\rm b}|-1)] C_{L-2}^{(0,0)}
    \nonumber \\
  + &\phi^{(0,0)}& \left( \begin{array}{c}
                        i^{\rm b}_1 \ldots i^{\rm b}_{r_{\rm b}-1}
                        i^{\rm b}_{r_{\rm b}+1} \ldots i^{\rm b}_{r_{\rm g}-1}
                        i^{\rm b}_{r_{\rm g}+1} \ldots i^{\rm b}_L \\
                        i^{\rm g}_1 \ldots i^{\rm g}_{r_{\rm b}-1}
                        i^{\rm g}_{r_{\rm b}+1} \ldots i^{\rm g}_{r_{\rm g}-1}
                        i^{\rm g}_{r_{\rm g}+1} \ldots i^{\rm g}_L \\
                       \end{array}
               \right)
\end{eqnarray}
is the desired mapping from the index to the number representation.

A possible configuration of the system for $(s_{\rm b},s_{\rm g}) = (1,1)$ is
illustrated in Fig.~\ref{Fig:lattice}, where the index representation of
the connectivity state for each completed row is shown to the right of the
figure.

\subsection{${\bf T}^{(2,0)}$ sector}

Considering now the case of two black strings, it appears that the
number of connectivity states for $L$ even is given by
\begin{equation}
  \tilde{C}_L^{(2,0)} = {L \choose 2} C_{L-2}^{(0,0)},
  \label{C_L(2,0)}
\end{equation}
where we have simply divided \Eq{C_L(1,1)} by 2 to take into account
the indistinguishability of two strings of the {\em same} flavour. This is
however not quite true, since for $L \ge 4$ the number (\ref{C_L(2,0)})
includes certain disallowed basis states. For $L=4$ these are
\begin{equation}
  \left( \begin{array}{rrrr} -1 & 1 & -1 & 1 \\ 0 & 0 & 0 & 0 \\ \end{array}
  \right) \mbox{ and }
  \left( \begin{array}{rrrr} 1 & -1 & 1 & -1 \\ 0 & 0 & 0 & 0 \\ \end{array}
  \right).
\end{equation}
The reason why these states are not valid is that, by definition of the
allowed vertices (see Fig.~\ref{Fig:vertices}), black loop segments
cannot cross a black string. In general, therefore, any configuration
where the positions of two equal, positive black indices are separated
by exactly one black string is not a valid one, even though the
positive indices of each flavour satisfy the well-nestedness criterion.
Accordingly, the true $C_L^{(2,0)}$ is less than the $\tilde{C}_L^{(2,0)}$
of \Eq{C_L(2,0)}.

We have not found it worthwhile to pursue the solution of this
complication, since the numbers $\tilde{C}_L^{(2,0)}$ are already less than
the $C_L^{(1,1)}$, and we need to diagonalise the transfer matrices
${\bf T}^{(1,1)}$ and ${\bf T}^{(2,0)}$ for the same values of $L$ in
order to determine the scaling dimensions $x_1$ and $x_2$ with the same
numerical precision. Instead we found it efficient to construct all the
$\tilde{C}_L^{(2,0)}$ basis states, list the number representations of
those that are disallowed, and force the corresponding entries of
${\bf T}^{(2,0)}$ to zero.

With this proviso the mapping from the index to the number representation
is
\begin{eqnarray}
  &\phi^{(2,0)}& \left( \begin{array}{c} {\bf i}^{\rm b} \\ {\bf i}^{\rm g}
                     \\ \end{array} \right)
  = [\psi({\bf i}^{\rm b} + {\bf 1}) - 1] C_{L-2}^{(0,0)} \nonumber \\
  + &\phi^{(0,0)}& \left( \begin{array}{c}
                        i^{\rm b}_1 \ldots i^{\rm b}_{r_1-1}
                        i^{\rm b}_{r_1+1} \ldots i^{\rm b}_{r_2-1}
                        i^{\rm b}_{r_2+1} \ldots i^{\rm b}_L \\
                        i^{\rm g}_1 \ldots i^{\rm g}_{r_1-1}
                        i^{\rm g}_{r_1+1} \ldots i^{\rm g}_{r_2-1}
                        i^{\rm g}_{r_2+1} \ldots i^{\rm g}_L \\
                       \end{array}
               \right),
\end{eqnarray}
where $r_1$ and $r_2$ are the positions of the two black strings, and
$\psi({\bf i}^{\rm b} + {\bf 1})$ means that we should lexicographically order
the positions of the black indices that are $-1$.

\end{multicols}

\end{document}